# Design optimization, commissioning, and uncertainty analysis of the quadrupole resonator system at Jefferson Lab for SRF material characterization

Mingqi Ge*,[1] Sarra Bira,[1] Kristof Brunner,[2] Natalie Gale,[1] Marco Garlaschè,[2] Valentin Giglia,[2] Oleksandr Hryhorenko,[1] Justin Kent,[1] Peter Owen,[1] Uttar Pudasaini,[1] Guillaume Rosaz,[2] Karol Scibor,[2] Pramita Tiwari,[1,3] Anne-Marie Valente-Feliciano,[1] Lorena Vega Cid,[2] Walter Venturini,[2] Haipeng Wang[1]

[1]Thomas Jefferson National Accelerator Facility, Newport News, Virginia, 23606, USA
[2]CERN, Esplanade des Particules 1, 1211 Geneva 23, Switzerland
[3]Old Dominion University, Norfolk, Virginia 23529, USA

**ABSTRACT**. A quadrupole resonator (QPR) provides a sample-based platform for characterizing materials for superconducting radio-frequency (SRF) applications under controlled field, frequency, and temperature conditions. This paper presents the design optimization, commissioning, and validation of the Jefferson Lab QPR system, including a quantitative assessment of measurement uncertainty. The resonator geometry was re-optimized from the CERN version-II design to improve quadrupole-mode separation and enable four usable modes at 400, 806, 1221, and 1640 MHz. The measurement system combines self-excited-loop RF operation, cable-loss-corrected power calibration, decay-based external-$Q$ calibration, and RF–DC thermal-substitution calorimetry to determine the peak surface magnetic field $B_{\mathrm{pk}}$ and sample surface resistance $R_{\mathrm{s}}$. Commissioning measurements on bulk Nb and $Nb_3Sn$-Ta-Cu samples validated the system response over a broad range of frequency, temperature, and RF field. The extracted superconducting energy-gap parameters are consistent with the reported values for Nb and $Nb_3Sn$, as well as with those obtained from single-cell cavity measurements. The commissioned system operates from 1.8 K to near the superconducting transition temperature of the sample, with accessible $B_{\mathrm{pk}}$ values from approximately 5 mT to a sample- and temperature-dependent heater-power-budget limit; a maximum field of 60 mT was demonstrated for bulk Nb at 400 MHz and 4 K. The combined relative standard uncertainties are 8.3% for $B_{\mathrm{pk}}$ and below 18% for $R_{\mathrm{s}}$ when $r = P_{\mathrm{DC2}}/P_{\mathrm{DC1}} < 0.9$. The worst-case $B_{\mathrm{pk}}$ resolution at the 95% confidence level is approximately 1.35 mT, while the $R_{\mathrm{s}}$ resolution is below 1 nΩ at 10 mT and 2 K. These results establish the JLab QPR as a calibrated, multi-frequency platform with quantified measurement uncertainty for SRF material characterization.

## I. INTRODUCTION.

Next-generation accelerators for fundamental research [1-3] and industrial applications [4, 5] impose increasingly demanding requirements on high-performance superconducting radio-frequency (SRF) cavities that sustain high accelerating gradients with low RF surface resistance, thereby reducing cryogenic load and overall system cost. These requirements have motivated extensive SRF materials research along two major directions. The first is to further improve the SRF performance of Nb cavities through advanced surface treatments and thin-film technologies [6-11], such as impurity doping [7, 8], mid-temperature baking [9] and high-power impulse magnetron sputtering (HiPIMS) [10, 11]. The second is to develop superconducting materials beyond conventional Nb, including $Nb_3Sn$, NbTiN, $MgB_2$, $V_3Sn$, and related higher-$T_c$ thin-film candidates [6,12], as well as advance thin-film structures such as superconductor-insulator-superconductor (SIS) multilayer structures [13, 14].

Among these research directions, $Nb_3Sn$ is particularly attractive because its higher critical temperature enables operation near 4 K with reduced cryogenic requirements, including the possibility of cryocooler-based conduction cooling, while its higher superheating field offers the potential for higher accelerating gradients [4, 5, 12]. Thin-film SRF cavity approaches are also attractive because they may reduce material cost and improve thermal stability compared with bulk-Nb cavities [6, 10, 11]. However, the RF performance of these materials cannot be fully assessed by room-temperature surface characterization techniques alone. Direct measurements of surface resistance under relevant RF

*Contact author: mingqi@jlab.org

field, frequencies, and cryogenic temperature conditions are required.

Single-cell cavity tests remain the most established method for evaluating superconducting material RF performance under cryogenic temperatures. However, they are relatively inefficient and costly for systematic materials screening. They usually provide measurements at a single operating frequency, require full cavity fabrication and processing, thus introducing intrinsic result variability arising from such processes. Post-test surface analysis is also difficult; in particular, regions associated with RF performance degradation, such as hot spots or localized lossy areas, cannot be re-examined nondestructively using detailed surface-analysis techniques.

These limitations motivate the use of sample-host resonators [15–17], in which small, reproducibly fabricated and replaceable samples can be tested under RF fields and subsequently examined using standard surface-analysis techniques. However, because the sample surface resistance is inferred from the total RF loss of the cavity, the measurement sensitivity of such systems can be limited by the background loss of the host cavity itself. This limitation becomes particularly important for advanced low-loss superconducting materials, such as $Nb_3Sn$, whose surface resistance can be substantially lower than that of Nb. Consequently, a sample-host resonator cannot reliably resolve the intrinsic surface resistance of a test material when it falls below the effective resolution set by the Nb cavity body [17].

The quadrupole resonator (QPR) was developed to address these limitations [18-20]. Instead of relying primarily on the quality factor of the host cavity, the QPR determines the sample surface resistance calorimetrically by comparing RF-induced heating with calibrated DC heating [18-20]. This approach reduces dependence on the RF losses of the resonator body and enables direct calorimetric determination of the power dissipated on the sample surface. In addition, the QPR can operate in multiple quadrupole modes, providing RF characterization at several discrete frequencies, typically from 400 to 1300 MHz, over temperatures from 2 K to near the sample's superconducting transition temperature, and at peak surface magnetic fields exceeding 100 mT depending on the implementation. QPR systems have been successfully developed and operated at CERN and HZB [18-21], demonstrating their value for SRF sample characterization. Recent sample-based studies, including samples removed from $Nb_3Sn$-coated cavities and Cu-based $Nb_3Sn$ samples prepared for QPR testing, further demonstrated the relevance of these sample-based workflows for advanced SRF materials development [22, 23].

The Jefferson Lab QPR is the first North American implementation of a quadrupole resonator system. This work presents its electromagnetic design optimization, system commissioning, RF and calorimetric measurement methodology, and experimental validation using representative Nb and $Nb_3Sn$-based samples. Comprehensive uncertainty and resolution analysis are also presented to define the practical measurement capabilities of the system.

## II. ELECTROMAGNETIC DESIGN AND GEOMETRY OPTIMIZATION

The Jefferson Lab quadrupole resonator (QPR) was designed [24] based on the CERN version-II concept, with the pole-shoe geometry re-optimized to increase the mode separation between the first quadrupole mode and the adjacent dipole mode. In the CERN version-II geometry, the separation between these modes was below 1 MHz [25], making mode identification and stable Low-level RF operation challenging during vertical testing. Accordingly, the JLab design optimization treated mode separation as a primary requirement, while keeping the remaining RF figures of merit within acceptable ranges to preserve the overall measurement performance of the QPR.

Table I. RF design requirements and optimized JLab QPR figures of merit for 1st QPR mode with bottom gap distance of 0.8 mm.

| Quantity | Unit | Requirement | Design Results |
|---|---|---|---|
| $\Delta f$ | MHz | $\geq 6$ | 10.60 |
| $\frac{B_{\mathrm{pk,samp}}}{B_{\mathrm{pk,cav}}}$ | | $\geq 0.86$ | 0.88 |
| $\frac{E_{\mathrm{pk,cav}}}{B_{\mathrm{pk,samp}}}$ | $\mathrm{MV\ m^{-1}\ mT^{-1}}$ | $\leq 0.21$ | 0.19 |
| $\frac{B_{\mathrm{pk,edge}}}{B_{\mathrm{pk,samp}}}$ | | $< 0.10$ | 0.04 |
| $c_1$ | $\mathrm{m^{-2}}$ | $> 1000$ | 1401 |
| $\frac{\int_{\mathrm{samp}} B^2\, dS}{U}$ | $\mathrm{mT^2 m^2\ J^{-1}}$ | $> 70$ | 89.70 |

The design objectives are summarized in Table I. The target frequency separation was set to at least 6 MHz between the first quadrupole mode near 400 MHz and the closest dipole mode. Additional constraints were imposed on the peak surface magnetic-field ratio between sample and cavity, the cavity peak surface electric field normalized to the sample peak magnetic field, the edge-to-sample peak magnetic-field ratio, the magnetic participation of the sample, and the constant $c_1 = B_{\mathrm{pk,samp}}^2 / \iint_{\mathrm{samp}} |B|^2 dS$, defined as the ratio of $B_{\mathrm{pk,samp}}^2$ to the surface integral of $B^2$ over the sample. These metrics were selected to measure the concentration of the magnetic field on the sample while limiting electric-field enhancement and RF leakage associated with edge heating.

The parameterized electromagnetic model used for optimization is shown in FIG 1. The independent geometric variables were the rod radius $R_{\text{rod}}$, loop height $h_{\text{loop}}$, loop radius $R_{\text{loop}}$, loop width $W_{\text{loop}}$, coil radius $R_{\text{coil}}$, loop separation $d_{\text{loop}}$, and the transition lengths $L_1$ and $L_2$.

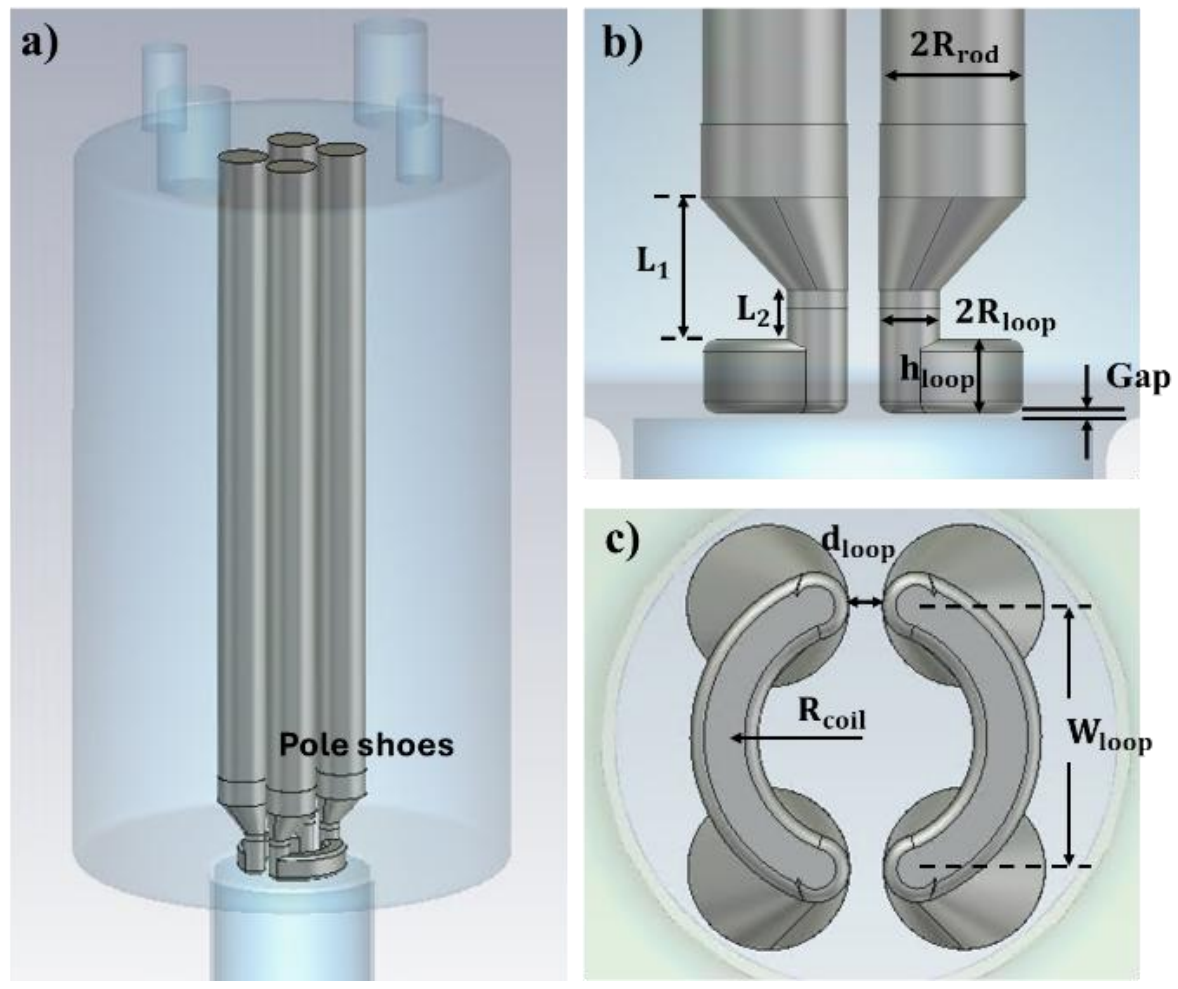

FIG 1. Parameterized electromagnetic model of the JLab QPR. (a) Full resonator geometry showing the four rods, pole shoes, and sample region; (b) side view of the rod-to-loop transition and gap; (c) top view of the loop geometry. The labeled dimensions define the variables scanned in the CST optimization.

The rod geometry controls both electric-field enhancement and modal separation. Increasing the rod radius $R_{\text{rod}}$ reduces the $E_{\text{pk,cav}}/B_{\text{pk,samp}}$ ratio because the RF charge is distributed over a larger conducting surface, while the $B_{\text{pk,samp}}/B_{\text{pk,cav}}$ ratio changes only weakly over the scanned range. The loop width $W_{\text{loop}}$ has a stronger influence on mode separation; the calculated sensitivity of $\Delta f$ was approximately 1 MHz/mm. A $R_{\text{rod}}$ of 12 mm was selected as a compromise among RF performance, cooling capacity and mechanical stability.

The loop and pole-shoe geometries primarily determine the field profile on the sample. The simulated magnetic field follows the pole-shoe shape and is concentrated near the sample center, which is desirable for the RF-DC thermal-substitution calorimetry because the RF dissipation should match the DC-heating profile. Smaller loop dimensions improve the field concentration, whereas larger loop width $W_{\text{loop}}$ and radius $R_{\text{rod}}$ improve mode separation and reduce field-emission risk. The adopted JLab geometry uses $R_{\text{loop}} = 5$ mm and $W_{\text{loop}} = 39$ mm to balance these competing requirements. Table II summarizes the JLab QPR geometry parameters.

The optimized geometry satisfies the RF design requirements in Table I. The closest quadrupole-dipole separation is increased to 10.6 MHz, exceeding the 6 MHz requirement and eliminating the severe mode-overlap limitation of the CERN version-II geometry. The optimized model also maintains a high sample-field ratio $B_{\text{pk,samp}}/B_{\text{pk,cav}}$, reduces the electric-field ratio $E_{\text{pk,cav}}/B_{\text{pk,samp}}$, and suppresses the magnetic field ratio $B_{\text{pk,edge}}/B_{\text{pk,samp}}$ at the 37.5-mm sample-edge radius.

Table II. Optimized geometric parameters of the JLab QPR. All dimensions are in millimeters.

| Parameter | Value (mm) |
|---|---|
| $R_{\text{rod}}$ | 12 |
| $h_{\text{loop}}$ | 8 |
| $R_{\text{loop}}$ | 5 |
| $W_{\text{loop}}$ | 39 |
| $R_{\text{coil}}$ | 20.9 |
| $d_{\text{loop}}$ | 15 |
| $L_1$ | 15 |
| $L_2$ | 8 |

Beyond resolving the mode-overlap problem, the optimized JLab geometry also expands the practical measurement frequency range relative to the CERN QPR. The CERN system focuses on the first three quadrupole modes up to ~1.2 GHz, whereas the JLab design supports four usable quadrupole modes at 400, 806, 1221, and 1640 MHz, shown in FIG 2. This fourth mode extends the upper frequency reach to 1640 MHz and increases the range over which frequency-dependent SRF surface resistance can be characterized using the same sample platform.

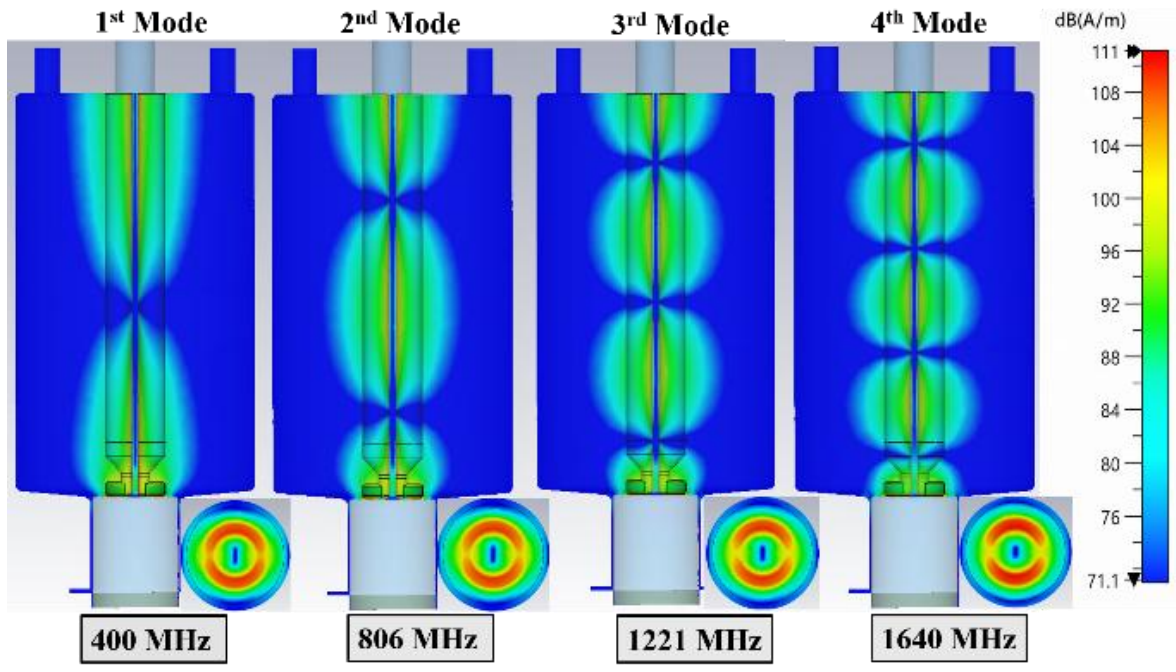

FIG 2. Simulated magnetic-field distributions of the four JLab QPR quadrupole modes at 400, 806, 1221, and 1640 MHz. The added 4th quadrupole mode extends the highest testing frequency up to 1640 MHz.

## III. COMMISSIONING

### A. QPR System at JLab

The Jefferson Lab QPR cavity body was fabricated from Nb at CERN in-house Workshops, using the optimized geometry described in Sec. II. To facilitate fabrication, the cavity body consists primarily of upper and lower sections, which are joined by a large stainless-steel ConFlat flange at the middle of the resonator. After fabrication, the cavity received approximately 200 μm of material removal by buffered chemical polishing (BCP). After shipment to Jefferson Lab, the cavity body was assembled and then high-pressure rinsed (HPR) through the coupler ports in a Class 10 cleanroom to minimize particulate contamination prior to vertical testing. FIG 3(a) shows the JLab QPR mounted on the test stand after clean assembly, with the RF couplers and sample installed. During sample installation, the replaceable sample was positioned from below the resonator using a UHV-compatible piezo platform, as shown in FIG 3(b). This configuration provided precise vertical control of the sample motion toward the sample port while maintaining cleanroom handling conditions and reducing the risk of particulate generation near the RF surface.

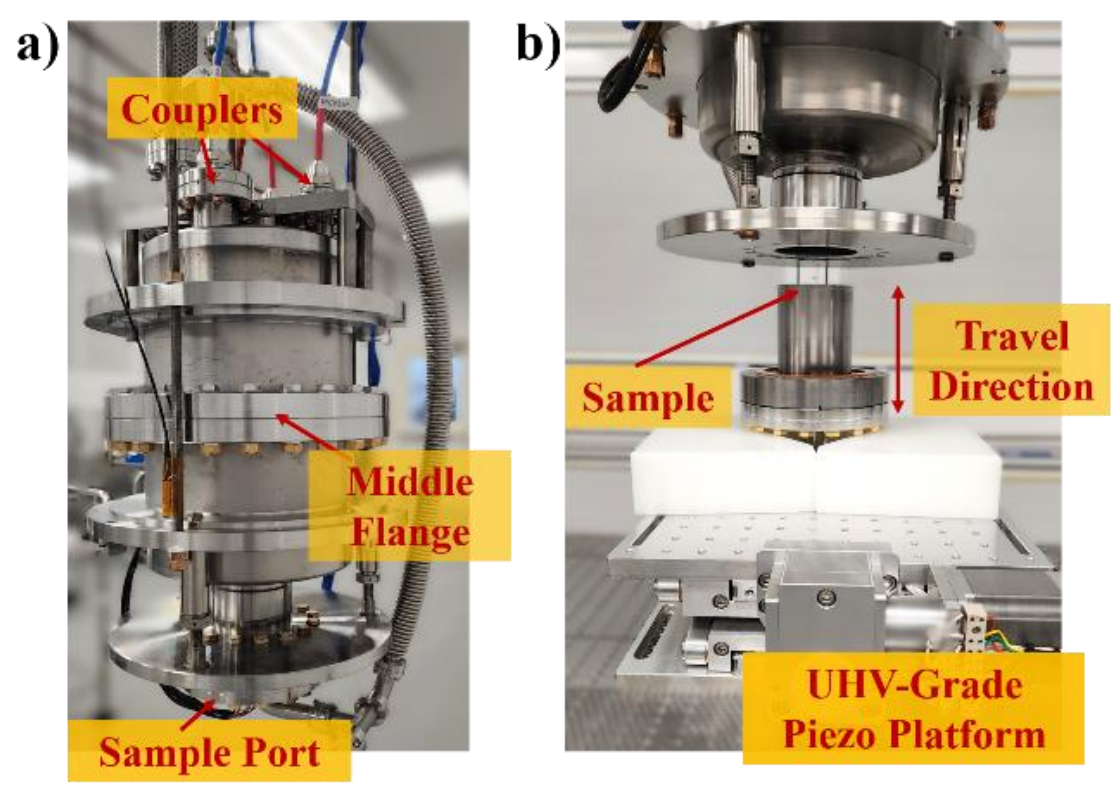

FIG 3. JLab QPR hardware and sample assembly in the cleanroom. (a) Fabricated JLab quadrupole resonator after chemical processing, high-pressure rinsing, and clean assembly, showing the RF couplers, middle flange, and sample port. (b) Sample installation using a UHV-grade piezo platform to provide controlled vertical positioning of the replaceable sample during cleanroom assembly.

#### *1. Instrumentation*

The sample-chamber instrumentation was designed to support RF–DC thermal-substitution calorimetry, as illustrated in FIG 4(a). A 50 Ω heater was mounted at the center of the backside of the sample to provide calibrated DC heating. A fluxgate magnetometer was installed near the sample to monitor the ambient magnetic-field level during cooldown, particularly as the sample passed through its superconducting transition. Multiple Cernox temperature sensors (A-C) were placed near the sample region, with their locations selected based on the simulated RF magnetic-field distribution on the sample surface, as shown in FIG 4(b). This placement ensured that the temperature readout was sensitive to the region where RF dissipation was expected to be large, and Cernox sensor D was mounted on the sample sidewall to monitor its temperature. In addition, a coil was installed to apply a controlled magnetic field during cooldown for flux-trapping studies.

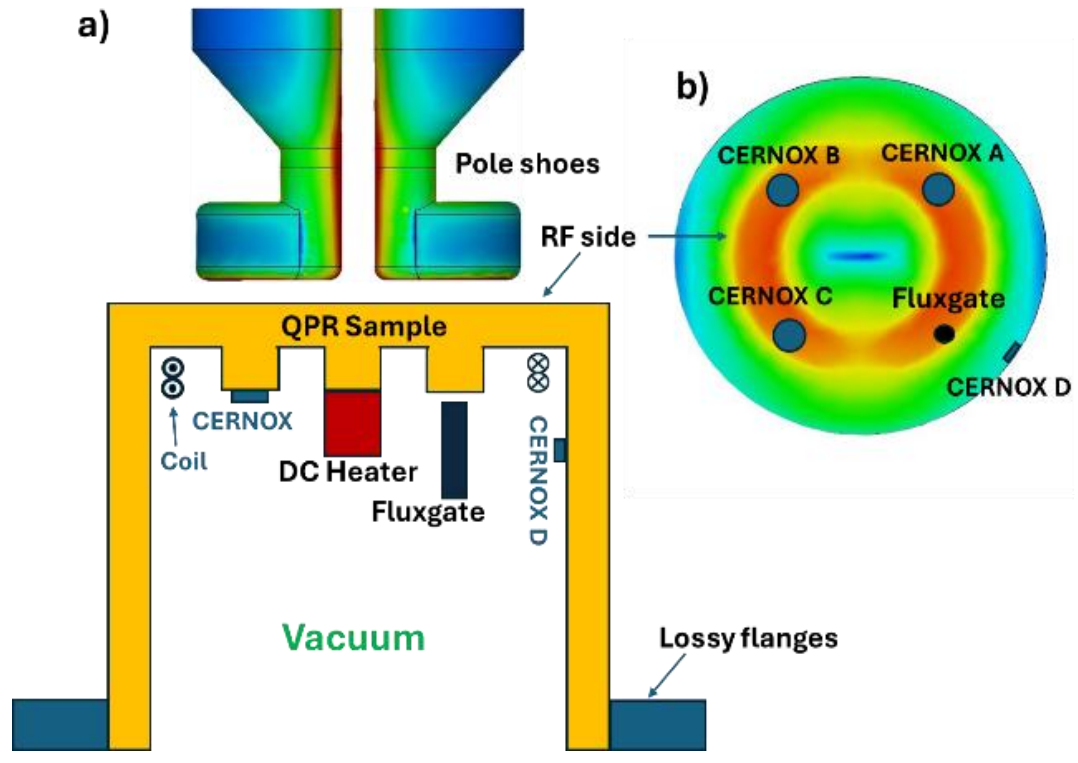

FIG 4. Instrumentation of the JLab QPR sample chamber. (a) Schematic layout of the internal sample-chamber instrumentation used for heater control, temperature readout, and magnetic-field monitoring. (b) Cernox sensor and fluxgate locations relative to the simulated RF magnetic-field distribution on the sample surface.

#### *2. LLRF and RF System*

In the JLab QPR RF test, the selected quadrupole mode was driven using the self-excited-loop (SEL) configuration shown in FIG 5 [26, 27]. The pickup signal from the resonator was down converted to a first intermediate frequency (IF) of 1.8 GHz and then to a second IF of 70 MHz, where it served as the feedback reference for phase and amplitude adjustment of the drive signal. The drive signal followed the reverse conversion path and was subsequently amplified and fed into the cavity through the input coupler. The feedback loop therefore tracked changes in the instantaneous resonant frequency of the selected QPR mode rather than operating at a fixed external frequency. The relevant RF power signals were measured by power meters and converted to cavity-reference-plane values using the corresponding cable-loss correction factors before further analysis.

The RF power-flow notation is defined in FIG. 6. In the standard two-coupler configuration, the resonator is driven through the input port and monitored through the pickup port. Because the four quadrupole modes have substantially different intrinsic quality factors, a third coupler was incorporated to provide additional flexibility for mode-dependent coupling optimization. In the three-coupler configuration, this additional coupler serves as a transmission port and introduces additional external loading, thereby modifying the effective input coupling condition. The calibrated forward and reflected powers at the input port are denoted by $P_{\mathrm{F}}$ and $P_{\mathrm{R}}$, respectively, while $P_{\mathrm{k}}$ and $P_{\mathrm{T}}$ denote the powers extracted through the pickup and transmission ports. The remaining RF dissipation consists of the cavity-wall loss $P_{\mathrm{c}}$ and the sample loss $P_{\mathrm{samp}}^{\mathrm{RF}}$.

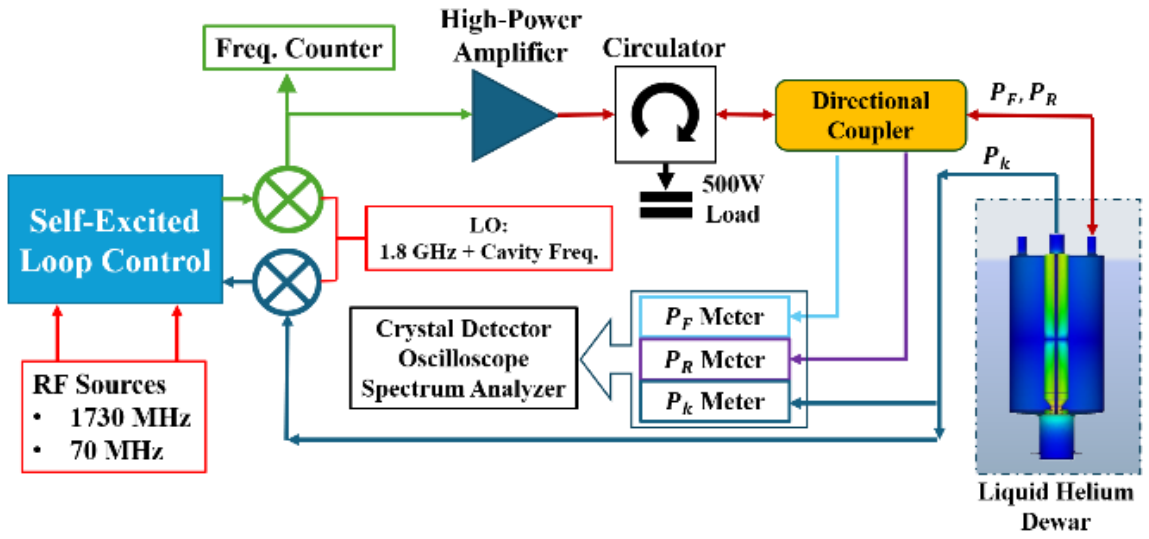

FIG 5. Illustration of LLRF and RF system for QPR tests at JLab.

For a stored energy $U$ and angular frequency $\omega_0$, the required forward power can be written as

$$P_{\mathrm{F}} = \left[\frac{(1+\beta_{\mathrm{in}}^*)^2}{4\beta_{\mathrm{in}}^*}\right]\left[\frac{\omega_0 U}{Q_0} + \frac{\omega_0 U}{Q_{\mathrm{k}}} + \frac{\omega_0 U}{Q_{\mathrm{T}}}\right], \quad (1)$$

where $Q_0$ is the intrinsic quality factor, $Q_{\mathrm{k}}$ and $Q_{\mathrm{T}}$ are the external quality factors of the pickup and transmission ports, respectively. $\beta_{\mathrm{in}}^*$ is the effective input coupling factor. The two-coupler case is recovered by setting $\frac{1}{Q_{\mathrm{T}}} = 0$. The definition of $\beta_{\mathrm{in}}^*$ and the derivation of the two- and three-coupler expressions are given in Appendix A.

FIG 7 compares the calculated forward power $P_{\mathrm{F}}$ as a function of the effective input coupling factor $\beta_{\mathrm{in}}^*$ for the two-coupler and three-coupler configurations. To estimate the RF power required at high field, the calculation was performed at a projected $B_{\mathrm{pk}} = 120$ mT on sample and using the measured low-field $Q_0$ values at 2 K: $1.5 \times 10^8$, $4.0 \times 10^6$, $1.4 \times 10^8$, and $3.5 \times 10^6$ for the first through fourth quadrupole modes, respectively. The lower values of the second and fourth quadrupole modes are attributed to additional RF loss at the middle flange, consistent with the field distributions shown in Fig. 2. The two-coupler configuration exhibits the expected minimum in $P_{\mathrm{F}}$ near effective critical coupling. For the three-coupler configuration, $Q_T = Q_{\mathrm{in}}$ was chosen to reduce the reflected power $P_{\mathrm{R}}$. Under this condition, the effective input coupling is constrained to $\beta_{in}^* < 1$, corresponding to a below-critical-coupling condition. However, this port also extracts RF power from the resonator; consequently, the reduction in reflected power is accompanied by an increased forward-power requirement. During commissioning, the two-coupler configuration was adopted, and the coupler associated with the additional port was adjusted to improve the coupling match of the second and fourth modes and thereby reduce the measurement uncertainty, as described in Sec. IV.

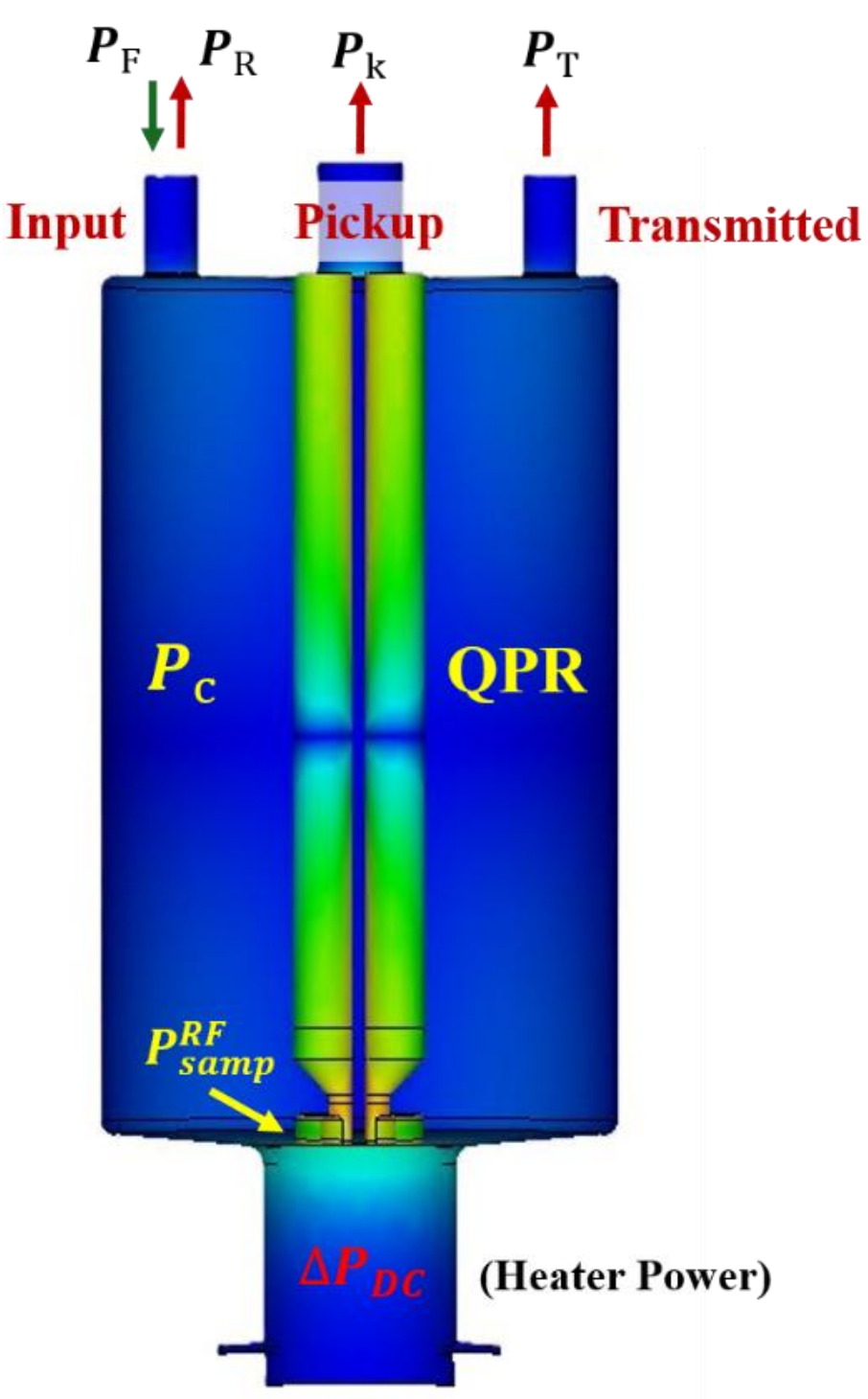

FIG 6. RF power-flow configuration for the JLab QPR test. $P_{\mathrm{F}}$ and $P_{\mathrm{R}}$ denote the forward and reflected powers, $P_{\mathrm{k}}$ and $P_{\mathrm{T}}$ denote the pickup and transmitted powers.

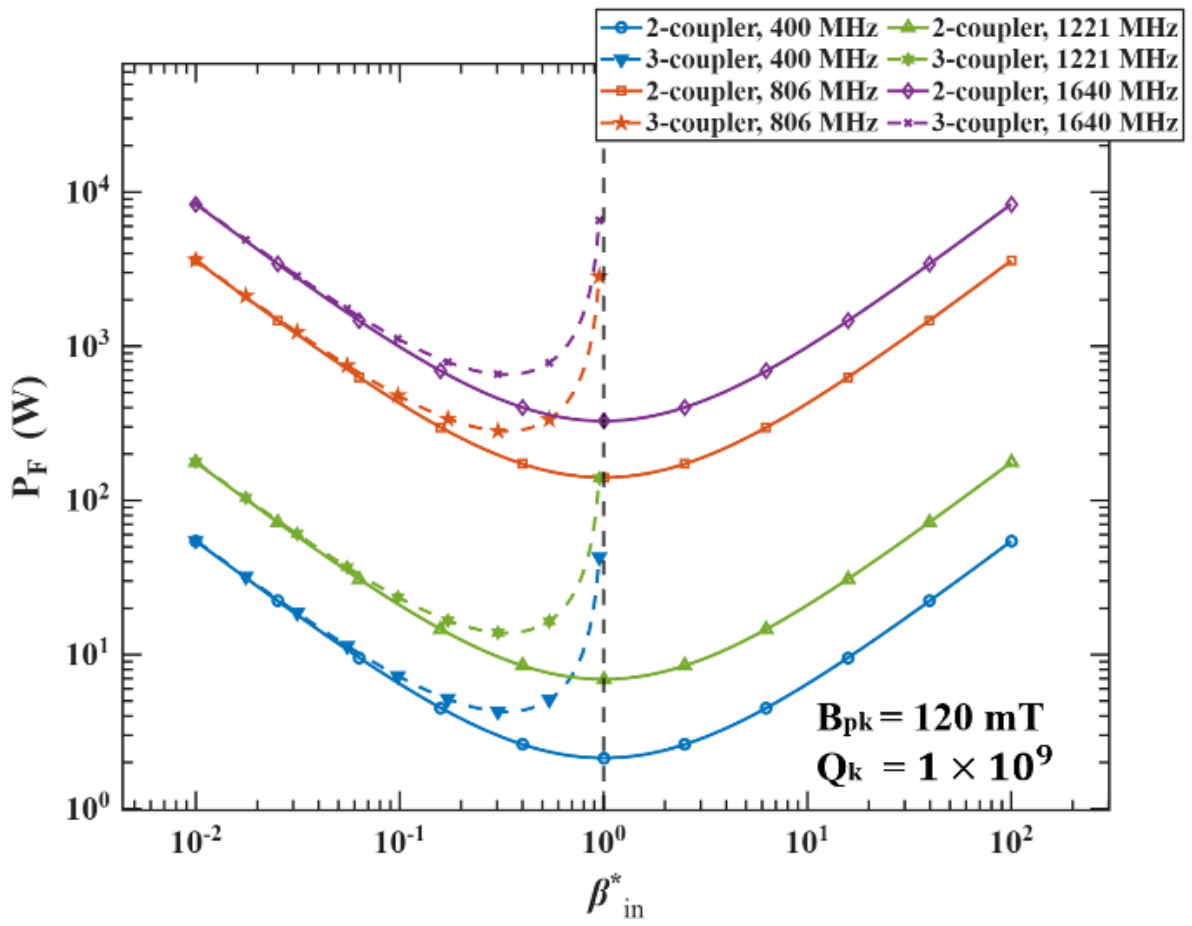

FIG 7. Calculated forward power $P_F$ versus effective input coupling factor $\beta_{in}^*$ for two-coupler and three-coupler QPR configurations at $B_{\mathrm{pk}} = 120\,\mathrm{mT}$, with $c_2 = B_{\mathrm{pk,samp}}^2/U = 0.13\,T^2\mathrm{J}^{-1}$ and $Q_{\mathrm{k}} = 1 \times 10^9$ and measured $Q_0$ is $1.5 \times 10^8$, $4 \times 10^6$, $1.4 \times 10^8$, $3.5 \times 10^6$ for 1st to 4th quadrupole modes respectively. For the three-coupler case, $Q_{\mathrm{T}} = Q_{\mathrm{in}}$, which constrains $0 < \beta_{\mathrm{in}}^* < 1$. The vertical dashed line marks $\beta_{\mathrm{in}}^* = 1$.

## B. Measurement and Results

### 1. $B_{\mathrm{pk}}$ and $R_{\mathrm{s}}$ Measurements

The peak surface magnetic field $B_{\mathrm{pk}}$ and surface resistance $R_{\mathrm{s}}$ on sample are computed from the Eq. (2) and (3) respectively as is shown,

$$B_{\mathrm{pk}} = \sqrt{\frac{c_2\, Q_{\mathrm{k}}\, P_{\mathrm{k}}}{\omega_0}}, \tag{2}$$

$$R_{\mathrm{s}} = 2 c_1 \mu_0^2 \frac{\Delta P_{DC}}{B_{\mathrm{pk}}^2}, \tag{3}$$

where $\omega_0$ is resonance angular frequency of QPR, $c_1$ and $c_2 = B_{\mathrm{pk,samp}}^2/U$ as constants can be obtained from CST Microwave Studio simulation, and Eq. (3) assumes a spatially uniform surface resistance over the sample. Because $c_1$ and $c_2$ depend on the sample-to-pole-shoe gap, the effective gap for each sample was inferred from the measured resonant frequency using the CST frequency-gap calibration, and the corresponding simulated $c_1$ and $c_2$ values were used in the analysis. The QPR measurement has two parts: (1) One-time decay measurement to determine $Q_{\mathrm{k}}$, and (2) the sample surface resistance is inferred via RF-DC thermal substitution, as is shown in Eq. (4).

$$\Delta P_{\mathrm{DC}} = P_{\mathrm{DC1}} - P_{\mathrm{DC2}} = I_{\mathrm{DC1}} V_{\mathrm{DC1}} - I_{\mathrm{DC2}} V_{\mathrm{DC2}} \tag{4}$$

where $\Delta P_{\mathrm{DC}}$ is the heater-power decrement required to reproduce the same measured thermal state, as indicated by the temperature-sensor readings, under RF-on and RF-off conditions. In practice, Cernox A was used as the temperature-regulation channel, while Cernox B provided an independent temperature readback and agreed with Cernox A at the millikelvin level. As illustrated in FIG 8(a) and (b), which show the temperature and heater-power responses, respectively, the sample is first heated to the target temperature using the DC heater, and the corresponding heater power is recorded as $P_{\mathrm{DC1}}$. The RF field is then turned on, and the temperature controller automatically reduces the heater power through the PID feedback loop to maintain the same target temperature. After thermal equilibrium is reached, the reduced heater power is recorded as $P_{\mathrm{DC2}}$, giving $\Delta P_{\mathrm{DC}} = P_{\mathrm{DC1}} - P_{\mathrm{DC2}}$.

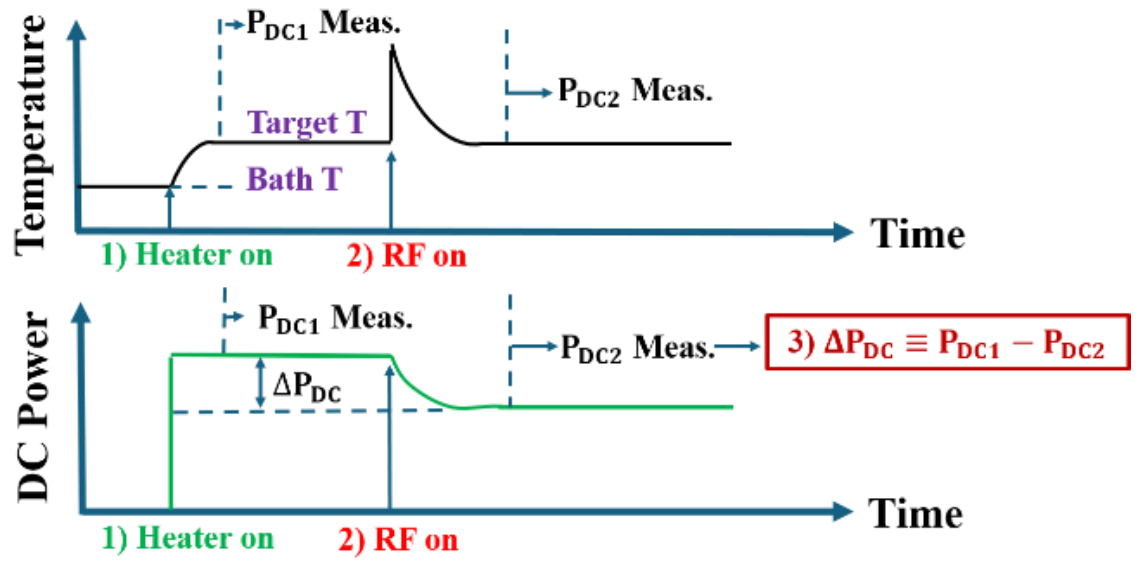

FIG 8. Schematic measurement sequence of the RF–DC substitution method used for QPR calorimetric surface-resistance measurements.

The maximum RF field accessible to the calorimetric measurement can be derived directly from Eq. (3). For a given bath temperature and target sample temperature, the available heater-power compensation range is limited by $P_{\mathrm{DC1}}$, since the heater power can only be reduced from its initial RF-off value ($P_{\mathrm{DC1}}$) to approximately zero after the RF field is applied. Therefore, by taking the maximum compensable RF heating as $\Delta P_{\mathrm{DC,max}} = P_{\mathrm{DC1}}$, Eq. (3) gives

$$B_{\mathrm{pk,max}} = \sqrt{\frac{2 c_1 \mu_0^2 P_{\mathrm{DC1}}}{R_{\mathrm{s}}}}, \tag{5}$$

FIG 9 defines the practical measurement boundary of the JLab QPR imposed by the heater-power budget and the expected sample surface resistance. The maximum measurable peak magnetic field, $B_{\mathrm{pk,max}}$, was calculated for Nb and $Nb_3Sn$-Ta-Cu samples at multiple bath temperatures: 2.5, 3, 4, 4.5K. For a given DC heater power $P_{\mathrm{DC1}}$, the RF power dissipated on the sample increases with $R_{\mathrm{s}} B_{\mathrm{pk}}^2$, therefore, $B_{\mathrm{pk,max}}$ decreases approximately as $R_{\mathrm{s}}^{-1/2}$. The results show that low-$R_s$ samples can be measured to substantially higher fields, while high-$R_{\mathrm{s}}$ samples are limited to lower $B_{\mathrm{pk}}$ by the available heater-power compensation range. The vertical lines mark the estimated superheating fields of Nb and $Nb_3Sn$, providing theoretical reference material limits. The heater-power budget is strongly affected by the overall thermal conductance of the sample assembly, including the substrate material, geometry, interfaces, and mounting conditions. At 4 K, oxygen-free high-conductivity (OFHC) copper has a representative thermal conductivity of approximately $6.4 \times 10^2$ - $1.9 \times 10^3$ $\mathrm{Wm^{-1}K^{-1}}$; whereas SRF-grade Nb with RRR ~300 has ~75 $\mathrm{Wm^{-1}K^{-1}}$ at 4.2 K. Thus, the thermal conductivity of the Cu substrate is approximately 10–25 times higher than that of Nb under comparable cryogenic conditions [28, 29]. Therefore, the Nb substrate requires a larger DC heater power than the Cu-based sample to reach the same target temperature, which provides a larger available RF-substitution power budget. This is reflected in FIG 9, where the Nb curves provide a higher $B_{\mathrm{pk,max}}$ envelope at the same assumed $R_{\mathrm{s}}$.

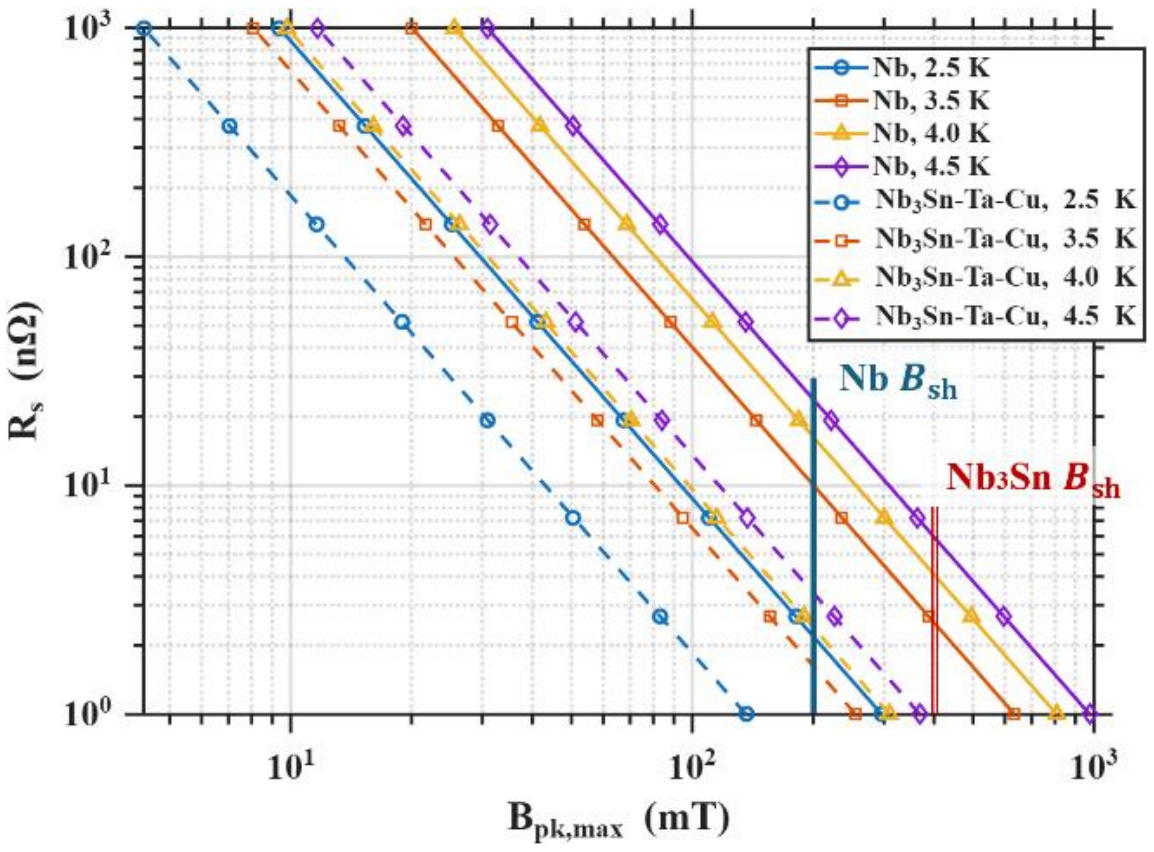

FIG 9. $B_{\mathrm{pk,max}}$ limit in the QPR measurements determined by $R_s$ and the heater power budget $P_{\mathrm{DC1}}$. Solid and dashed curves represent Nb and $Nb_3Sn$-Ta-Cu, respectively, and the vertical lines indicate the corresponding Nb and $Nb_3Sn$ superheating fields $B_{\mathrm{sh}}$.

### *2. Measurement Results*

Two samples made from different SRF materials were measured during commissioning: a bulk Nb sample treated by light electropolishing, denoted JN1, and a $Nb_3Sn$-Ta-Cu sample coated by magnetron sputtering, denoted P1, as shown in FIG 10. In the $Nb_3Sn$-Ta-Cu sample, Ta serves as the barrier layer between the $Nb_3Sn$ film and the copper substrate. The comparison of these two material samples provides an independent cross-check of the QPR measurement system.

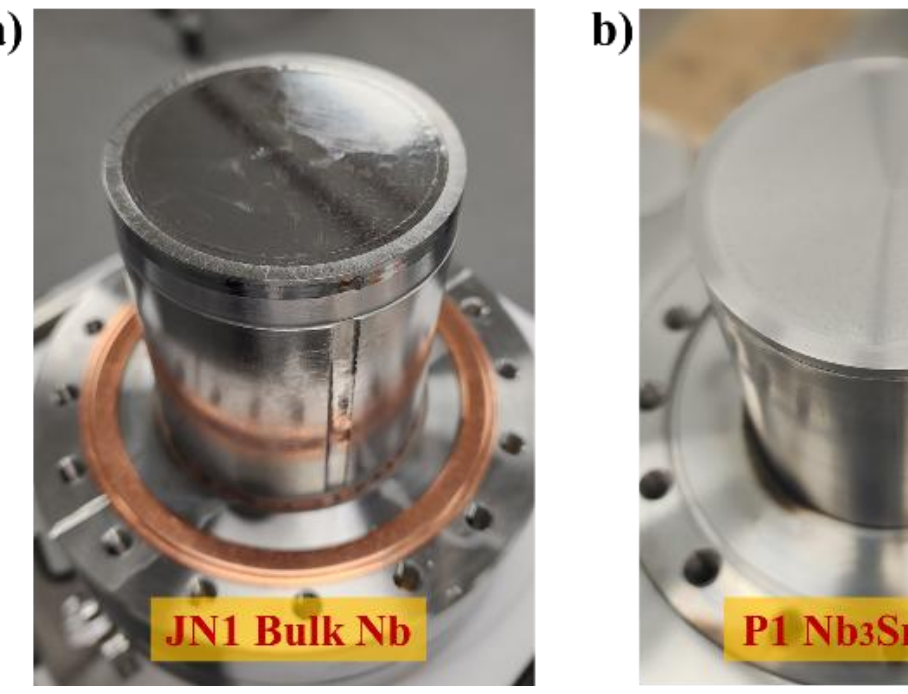

FIG 10. QPR sample (a) JN1 Bulk Nb; (b) $Nb_3Sn$-Ta-Cu.

During the JN1 tests, the liquid-helium bath temperature was 1.8 K. FIG 11 shows the field-dependent surface resistance of the Nb sample measured at 400, 806, and 1221 MHz. For each mode, $R_{\mathrm{s}}$ vs. $B_{\mathrm{pk}}$ was measured at sample temperatures of 2.5 and 4 K. The maximum accessible $B_{\mathrm{pk}}$ was limited by the heater-power-budget boundary described in FIG 9, rather than by quench. The highest field reached was 60 mT at 400 MHz and 4 K, corresponding to an equivalent accelerating gradient $E_{\mathrm{acc}}$ of approximately 14 MV/m for a TESLA-shaped cavity [30].

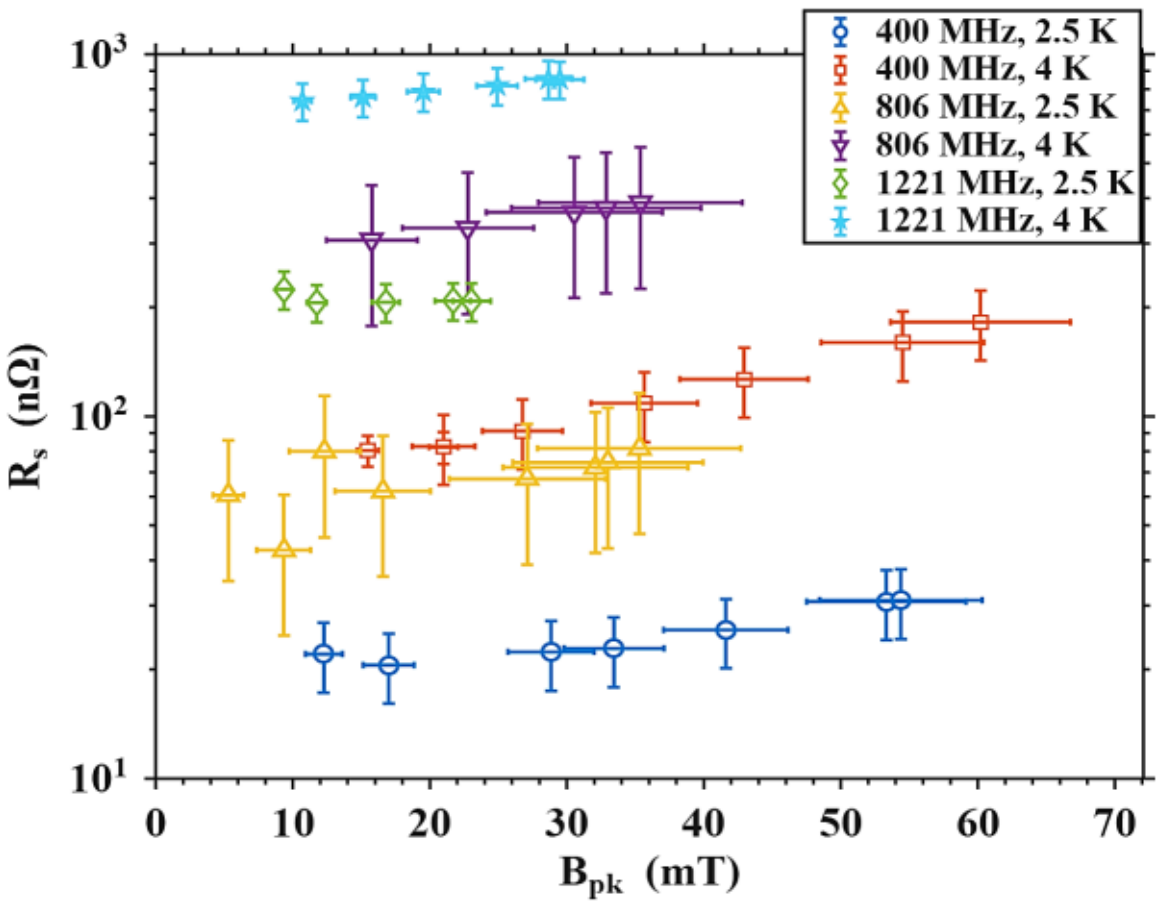

FIG 11. Field-dependent surface resistance of the Nb sample JN1 measured at multiple temperatures and frequencies.

FIG 12 shows the temperature-dependent surface resistance of the JN1 sample measured at 400, 806, and 1221 MHz. The lowest measured $R_{\mathrm{s}}$ was approximately 15 nΩ at 400 MHz and 1.8 K. The measurements covered temperatures from 1.8 K to approximately 4.5-5.0 K, providing the temperature range needed for comparison with the single-cell cavity results discussed in Sec. III.C.

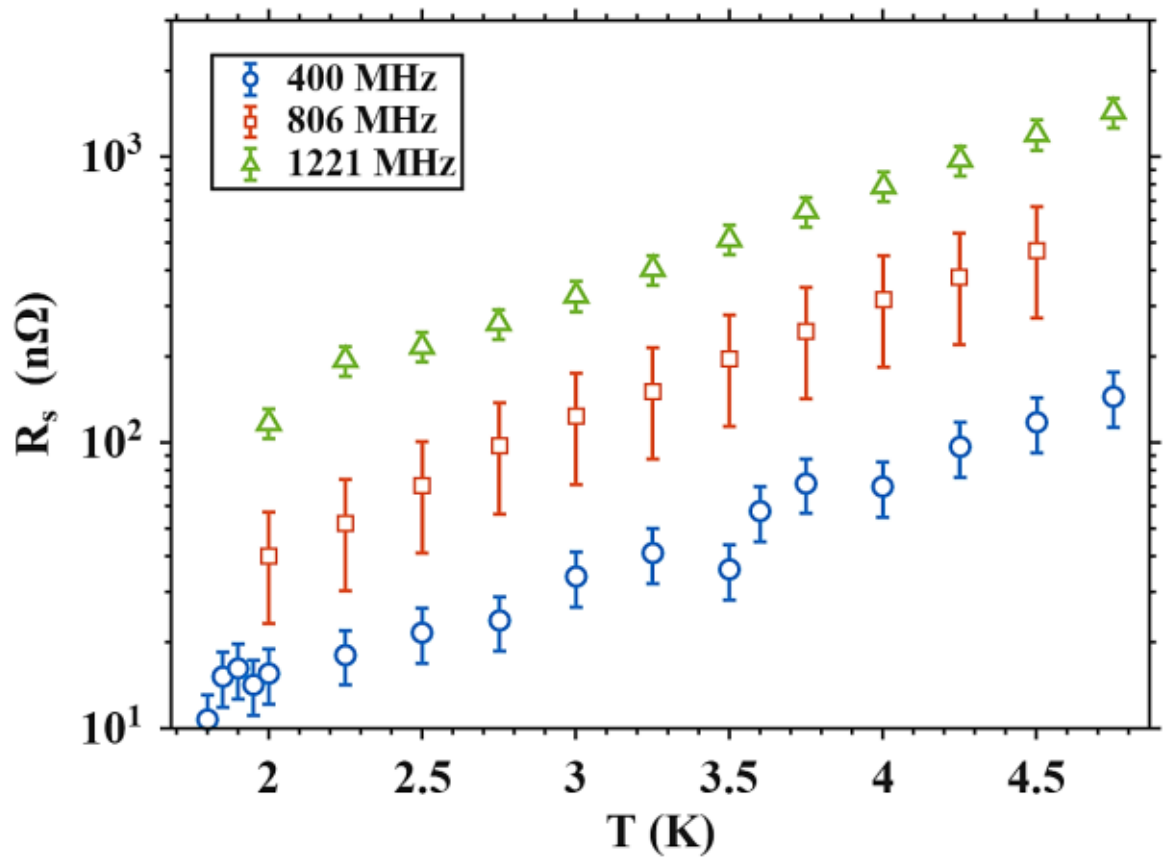

FIG 12. Temperature dependence of the surface resistance of the Nb sample JN1 measured at 400, 806, and 1221 MHz and with $B_{\rm pk} = 18$ mT.

FIG 13 shows the field-dependent surface resistance of the $Nb_3Sn$-Ta-Cu sample. The bath temperature was 2K, and measurements were obtained at 400, 806, 1221, and 1640 MHz, demonstrating operation of the JLab QPR over four quadrupole modes. Compared with the Nb sample, the $Nb_3Sn$-Ta-Cu sample exhibits lower $B_{\rm pk,max}$ (~25 mT) due to its smaller available heater-power budget.

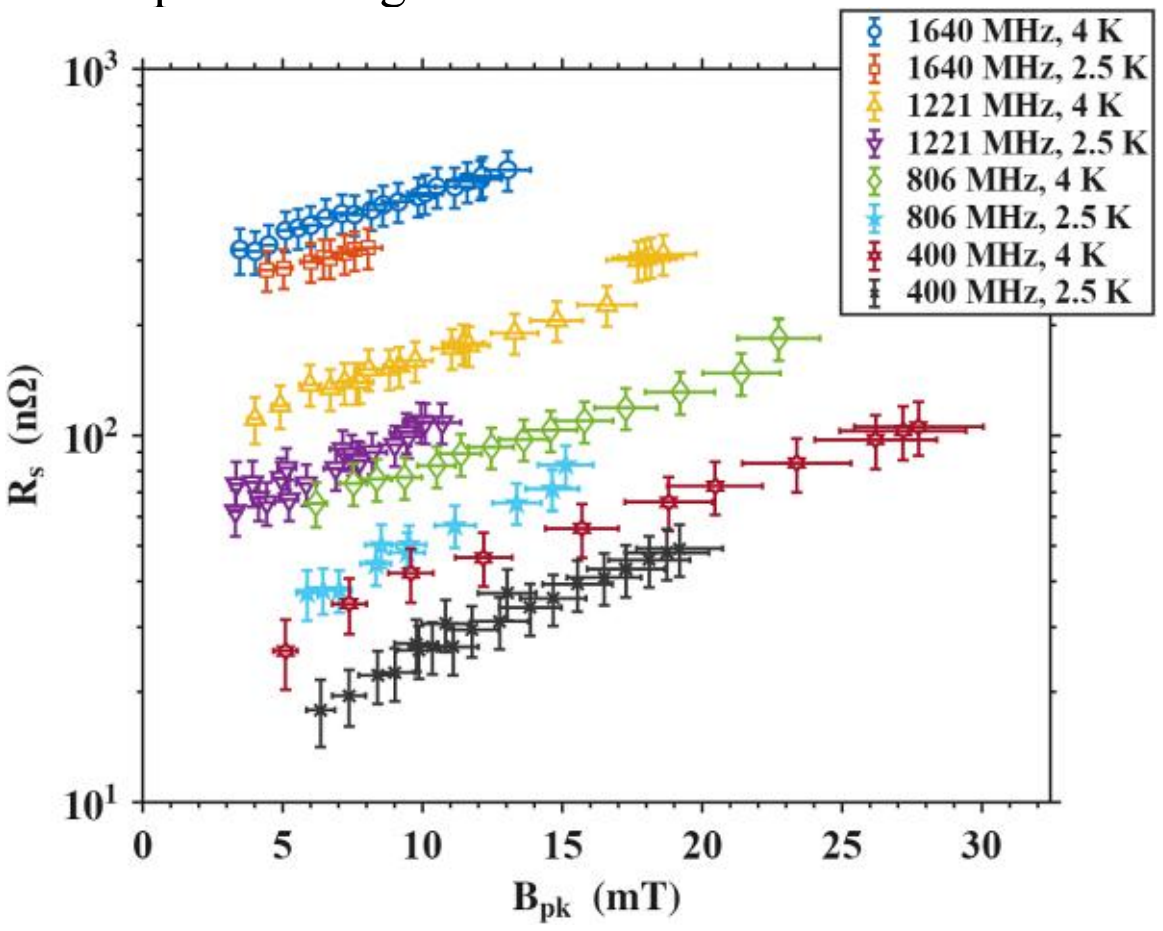

FIG 13. Field-dependent surface resistance of the $Nb_3Sn$-Ta-Cu sample P1 measured at multiple temperatures and frequencies.

The temperature-dependent results for the $Nb_3Sn$-Ta-Cu sample are shown in FIG 14. The measured $R_s$ increases rapidly at 4.0-5.5 K, particularly at the lower-frequency modes.

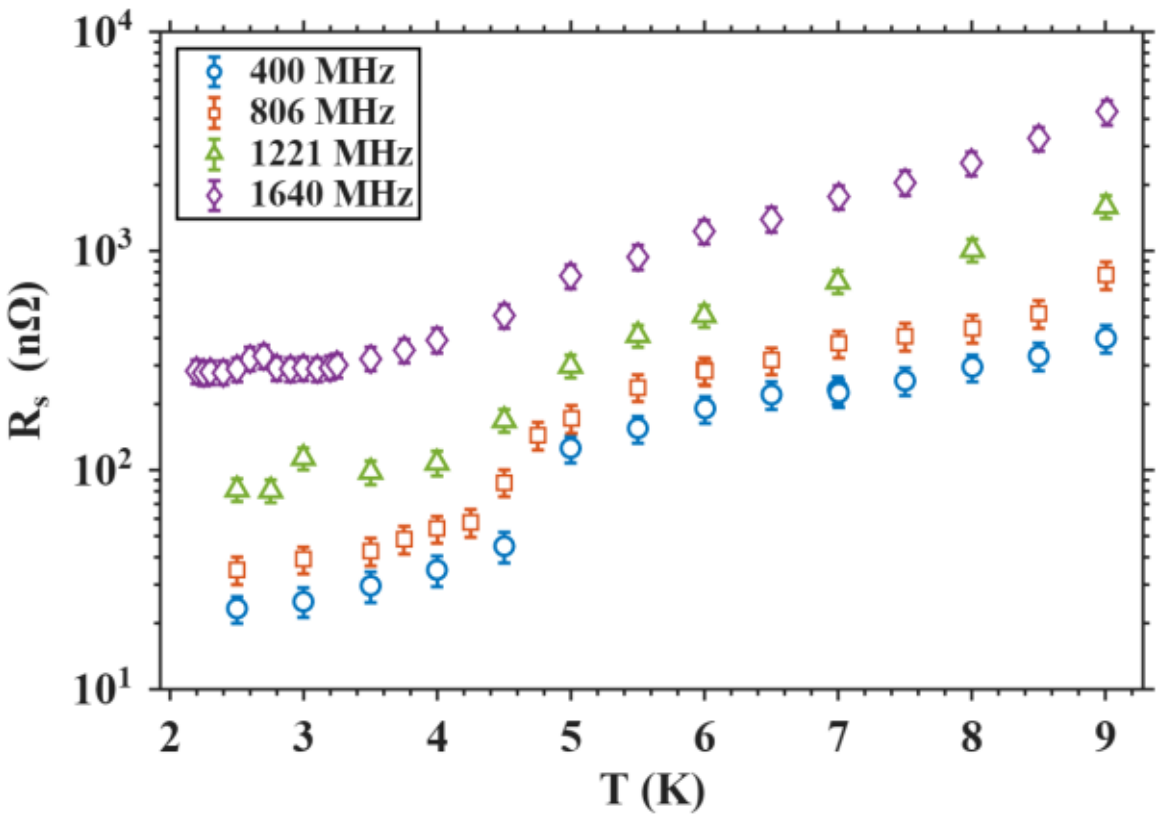

FIG 14. Temperature dependence of the surface resistance of the $Nb_3Sn$-Ta-Cu sample P1, measured at four frequencies—400, 806, 1221, and 1640 MHz—with $B_{\rm pk} = 10$ mT.

Taken together, FIGs. 11–14 summarize the main RF performance results from the commissioning measurements. The temperature-dependent datasets are further analyzed below using an Arrhenius-type representation to extract the effective superconducting energy gap and provide an additional validation of the measured surface resistance.

### C. Analysis and Discussion

To validate the measured results, the temperature-dependent surface resistances of Nb and $Nb_3Sn$ were analyzed using an Arrhenius-type method to extract the effective superconducting energy gaps. These extracted gap values provide useful benchmarks because Nb and $Nb_3Sn$ have been extensively studied and their superconducting properties are well established. The temperature dependent surface resistance can be written as Eq. (6) [31],

$$R_{\rm s} = A\frac{f^2}{T}\exp\left(-\frac{\Delta}{k_{\rm B}T}\right) + R_0 \tag{6}$$

where $R_0$ is the residual resistance obtained from the fit to Eq. (6) and subsequently fixed in the Arrhenius analysis, $\Delta/k_B$ is the effective superconducting energy gap, defining

$$Y = \ln\left(\frac{(R_s - R_0)T}{f^2}\right), \quad \text{and} \quad X = \frac{1}{T},$$

Eq. (6) can be rewritten as

$$Y = \ln A - \frac{\Delta}{k_B}X, \tag{7}$$

FIG 15 presents the Arrhenius analyses for the two samples measured during QPR commissioning. The electropolished bulk-Nb sample JN1 was measured at 400, 806, and 1221 MHz, while the magnetron-sputtered $Nb_3Sn$-Ta-Cu sample P1 was measured at 400, 806, 1221, and 1640 MHz. FIG 15(a) compares the JN1 results with those from the 1.3 GHz bulk-Nb single-cell cavities ML-02 and RDT-5 [32] treated by electropolishing and bi-polar electropolishing, respectively. FIG 15(b) compares the P1 results with those from the 1.3 GHz vapor-diffused $Nb_3Sn$-Nb single-cell cavity TE1NS001. The approximately linear behavior over the selected fitting ranges supports the Arrhenius approximation used in Eq. (7).

For the bulk-Nb sample JN1, the extracted $\Delta/k_B$ values are 16.0, 18.5, and 16.4 K at 400, 806, and 1221 MHz, respectively, giving a three-frequency mean of 17.0 K. The corresponding values obtained from the single-cell cavities ML-02 and RDT-5 are 19.2 and 18.3 K, respectively, giving a mean of 18.8 K. The mean of the three QPR-derived values is therefore 1.8 K, or 9.5%, lower than the mean of the two single-cell cavity values.

The commonly reported energy-gap parameter for Nb is approximately 15-18 K [31]. Calculations including electron-phonon strong-coupling effects give $\Delta/k_B \approx 18.0$ K, compared with the weak-coupling BCS value of approximately 16.6 K [33]. The QPR mean lies within this expected range, while the single-cell mean is slightly above the range but remains close to the strong-coupling value. Considering the different surface treatments and measurement configurations, the agreement between the QPR and single-cell results is reasonable.

For the magnetron-sputtered $Nb_3Sn$-Ta-Cu sample P1, the extracted $\Delta/k_B$ values are 39.9, 35.5, 34.2, and 39.3 K at 400, 806, 1221, and 1640 MHz, respectively, giving a four-frequency mean of 37.2 K. The vapor-diffused $Nb_3Sn$-Nb single-cell cavity TE1NS001 yields a value of 34.6 K; therefore, the mean of QPR-derived values is 2.6 K, or 7.6%, higher than the single-cell value.

$Nb_3Sn$ is a strong-coupling superconductor, with a reported gap ratio $2\Delta/(k_B T_c)$ of approximately 4.2–4.4 [34]. For stoichiometric $Nb_3Sn$ $T_c \approx 18$ K, this corresponds to $\Delta/k_B \approx 37.8$–$39.6$ K. The QPR mean is close to this expected range, while the single-cell result is moderately lower. Because P1 and TE1NS001 were produced by different coating processes on different substrates and were evaluated at different RF frequencies and measurement configurations, the observed difference is reasonable. The consistency of the QPR-extracted energy-gap values across frequencies for both Nb and $Nb_3Sn$ also supports the approximately $f^2$ dependence of $R_{\mathrm{BCS}}$ assumed in Eq. (6). The fit uncertainty of each extracted energy-gap parameter was obtained from the standard error of the Arrhenius-fit slope. The experimental uncertainty was propagated from the measured $R_s$ uncertainty using Eq. (6), and the median of the pointwise propagated uncertainties over the fitting range was taken as the representative experimental uncertainty. The extracted values, together with their fitting and experimental uncertainties, are summarized in Table III.

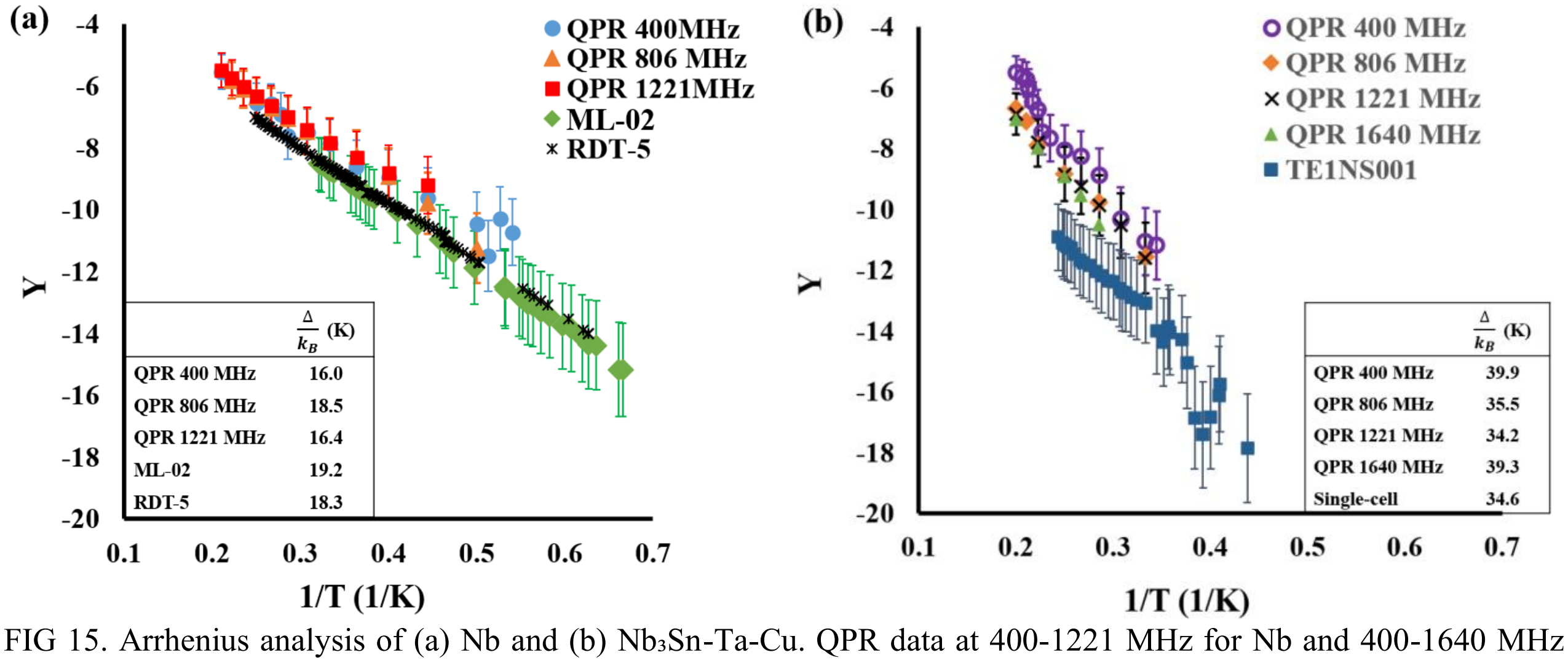

FIG 15. Arrhenius analysis of (a) Nb and (b) $Nb_3Sn$-Ta-Cu. QPR data at 400-1221 MHz for Nb and 400-1640 MHz for $Nb_3Sn$-Ta-Cu are compared with 1.3 GHz single-cell results from ML-02/RDT-5 and TE1NS001, respectively. Extracted energy-gap parameters are listed in Table III.

Table III. Comparison of superconducting energy-gap parameters obtained from QPR and single-cell cavity measurements.

| Cavity | Materials | Frequency (MHz) | Extracted $\Delta/k_B$ (K) | $\Delta/k_B$ fit uncertainty (%) | $\Delta/k_B$ experimental uncertainty (%) |
|---|---|---|---|---|---|
| QPR | Nb | 400 | 16.0 | 4.9 | 6.1 |
| QPR | Nb | 806 | 18.5 | 2.8 | 4.4 |
| QPR | Nb | 1221 | 16.4 | 3.9 | 5.4 |
| Single-cell (ML-02) | Nb | 1300 | 19.2 | 0.6 | 1.4 |
| Single-cell (RDT-5) | Nb | 1300 | 18.3 | 0.3 | 0.8 |
| QPR | $Nb_3Sn$-Ta-Cu | 400 | 39.9 | 5.0 | 4.8 |
| QPR | $Nb_3Sn$-Ta-Cu | 806 | 35.5 | 5.1 | 5.8 |
| QPR | $Nb_3Sn$-Ta-Cu | 1221 | 34.2 | 5.0 | 6.9 |
| QPR | $Nb_3Sn$-Ta-Cu | 1640 | 39.3 | 4.2 | 8.0 |
| Single-cell (TE1NS001) | $Nb_3Sn$-Nb | 1300 | 34.6 | 5.1 | 7.8 |

## IV. ERROR ANALYSIS

As described by Eqs. (2)-(4) in Sec. III B, the uncertainty in $B_{\mathrm{pk}}$ is mainly governed by the calibration uncertainty of $Q_k$, while the uncertainty in $R_s$ is affected by both the propagated uncertainty in $B_{\mathrm{pk}}$ and the uncertainty in $\Delta P_{\mathrm{DC}}$. In this section, standard uncertainty propagation is used to evaluate the corresponding error levels.

For each directly measured quantity $x$, the combined standard uncertainty is calculated as

$$u_c^2(x) = u_m^2(x) + u_{\mathrm{inst}}^2(x), \quad (8)$$

here: $u_m(x)$ is measurement repeatability and offset uncertainty (Type A) and $u_{\mathrm{inst}}(x)$ is standard uncertainty derived from the instrument specifications. (Type B). In this work, all measured inputs are assumed statistically independent. Therefore, covariance terms are neglected throughout the uncertainty analysis.

### A. Instrument Specifications

For the voltage and current measurements, the instrument-specification uncertainties summarized in Table IV were obtained from the Keithley 2000 specifications [35] and converted to standard uncertainties by assuming a rectangular distribution. The relative standard uncertainty adopted for the RF-power measurement is 7 %, which includes the uncertainties associated with the cable calibration, power meter, and power sensor head [36].

Table IV. Final instrument uncertainty values used in uncertainty evaluation

| | Instrument | Standard Uncertainty (1σ) |
|---|---|---|
| Voltage (V) | Keithley 2000 (10 V) | $\frac{20\times10^{-6}\mid V\mid+50\times10^{-6}}{\sqrt{3}}$ |
| Current (A) | Keithley 2000 (100 mA) | $\frac{300\times10^{-6}\mid I\mid+84\times10^{-6}}{\sqrt{3}}$ |

### B. One-time Calibration of $Q_{\mathrm{k}}$

$Q_{\mathrm{k}}$ is a critical parameter in QPR measurements. It is determined at a reference operating point $x_0$ (typically at low field) When the stored energy decays as $U(t) = U_0\exp(-t/\tau)$, the loaded quality factor is

$$Q_L = \omega_0\tau, \quad (9)$$

where $\tau$ is the decay time constant. As derived in Appendix B, $Q_{\mathrm{k}}$ can be written in closed form as

$$Q_k = \frac{2\sqrt{P_{\mathrm{F}}}(\sqrt{P_{\mathrm{F}}}\pm\sqrt{P_{\mathrm{R}}})\omega_0\tau}{P_{\mathrm{k}}}, \quad (10)$$

The sign choice ($+$ or $-$) corresponds to over-coupled and under-coupled condition of $\beta_{in}^*$, respectively. The one-time calibration of $Q_{\mathrm{k}}$ is treated as a first-stage uncertainty propagation problem evaluated at the reference operating point. Since the calibrated $Q_{\mathrm{k}}$ is subsequently used in the continuous evaluation of $B_{\mathrm{pk}}$ and $R_{\mathrm{s}}$, its uncertainty acts as a common, correlated calibration uncertainty in all downstream results. During temperature-dependent measurements, only the sample temperature is varied, while the QPR cavity remains at the helium-bath temperature; therefore, the cavity coupling condition remains essentially unchanged over the sample-temperature scan. Use the input vector

$$X_{Q_k} = [P_{\mathrm{F}}, P_{\mathrm{R}}, P_{\mathrm{k}}, \tau]^T, \tag{11}$$

the Jacobian is from Eq. (10)

$$J_{Q_\mathrm{k}} = \left[\frac{\partial Q_\mathrm{k}}{\partial P_\mathrm{F}}\ \frac{\partial Q_\mathrm{k}}{\partial P_\mathrm{R}}\ \frac{\partial Q_\mathrm{k}}{\partial P_\mathrm{k}}\ \frac{\partial Q_\mathrm{k}}{\partial \tau}\right] = \left[\frac{2\omega_0\tau}{P_\mathrm{k}}\left(1+\frac{s}{2}\sqrt{\frac{P_\mathrm{R}}{P_\mathrm{F}}}\right),\ \frac{2\omega_0\tau}{P_\mathrm{k}}\left(\frac{s}{2}\sqrt{\frac{P_\mathrm{F}}{P_\mathrm{R}}}\right),\ -\frac{Q_\mathrm{k}}{P_\mathrm{k}},\ \frac{Q_\mathrm{k}}{\tau}\right], \tag{12}$$

where $s = \pm 1$ with $+$ for over-coupling and $-$ for under-coupling, under the independent-input assumption, the uncertainty of any output $y = g(\mathbf{x})$ is propagated as

$$u_c^2(Q_\mathrm{k}) = J_{Q_\mathrm{k}} U_{X_{Q_\mathrm{k}}} J_{Q_\mathrm{k}}^T = \sum_i \left(\frac{\partial Q_\mathrm{k}}{\partial X_{Q_\mathrm{k},i}}\right)^2 u_c^2(X_{Q_\mathrm{k},i}), \tag{13}$$

and

$$\left(\frac{u_c(Q_\mathrm{k})}{Q_\mathrm{k}}\right)^2 = \left[\frac{2\sqrt{P_\mathrm{F}}+s\sqrt{P_\mathrm{R}}}{2\left(\sqrt{P_\mathrm{F}}+s\sqrt{P_\mathrm{R}}\right)}\right]^2\left(\frac{u_c(P_\mathrm{F})}{P_\mathrm{F}}\right)^2 + \left[\frac{\sqrt{P_\mathrm{R}}}{2\left(\sqrt{P_\mathrm{F}}+s\sqrt{P_\mathrm{R}}\right)}\right]^2\left(\frac{u_c(P_\mathrm{R})}{P_\mathrm{R}}\right)^2 + \left(\frac{u_c(P_\mathrm{k})}{P_\mathrm{k}}\right)^2 + \left(\frac{u_c(\tau)}{\tau}\right)^2 \tag{14}$$

Eq. (13) gives the combined standard uncertainty of the calibrated output, while Eq. (14) gives the corresponding relative combined standard uncertainty.

The relative uncertainty in $Q_\mathrm{k}$ is strongly affected by coupling factor $\beta_{\mathrm{in}}^*$, the Eq. (14) can be rewritten in the form of Eq. (15). This form is convenient for experimental interpretation because it separates the contribution associated with the decay measurement from that associated with the RF power terms:

$$\left(\frac{u_c(Q_\mathrm{k})}{Q_\mathrm{k}}\right)^2 = \left(\frac{3\beta_{\mathrm{in}}^*+1}{4\beta_{\mathrm{in}}^*}\right)^2\left(\frac{u_c(P_\mathrm{F})}{P_\mathrm{F}}\right)^2 + \left(\frac{\beta_{\mathrm{in}}^*-1}{4\beta_{\mathrm{in}}^*}\right)^2\left(\frac{u_c(P_\mathrm{R})}{P_\mathrm{R}}\right)^2 + \left(\frac{u_c(P_\mathrm{k})}{P_\mathrm{k}}\right)^2 + \left(\frac{\sqrt{2}\beta_{\mathrm{in}}^*(VSWR-1)}{(\beta_{\mathrm{in}}^*+1)(VSWR+1)}\right)^2, \tag{15}$$

with $\frac{u(\tau)}{\tau} \approx \frac{\sqrt{2}\beta_{\mathrm{in}}^*(VSWR-1)}{(\beta_{\mathrm{in}}^*+1)(VSWR+1)}$. [36, 37]

where VSWR is the voltage standing-wave ratio of the RF measurement system.

FIG 16 shows the relative uncertainty of $Q_\mathrm{k}$ as a function of $\beta_{\mathrm{in}}^*$. In the under-coupled regime, the first and second terms in Eq. (15) dominate the uncertainty budget. As $\beta_{\mathrm{in}}^*$ decreases and approaches 0, the contributions from these two terms increase rapidly, leading to a substantial growth in the overall uncertainty of the calibrated $Q_\mathrm{k}$. In contrast, in the overcoupled regime, the fourth term becomes the dominant contribution. For the JLab LLRF system, the typical VSWR is approximately 1.10-1.20.

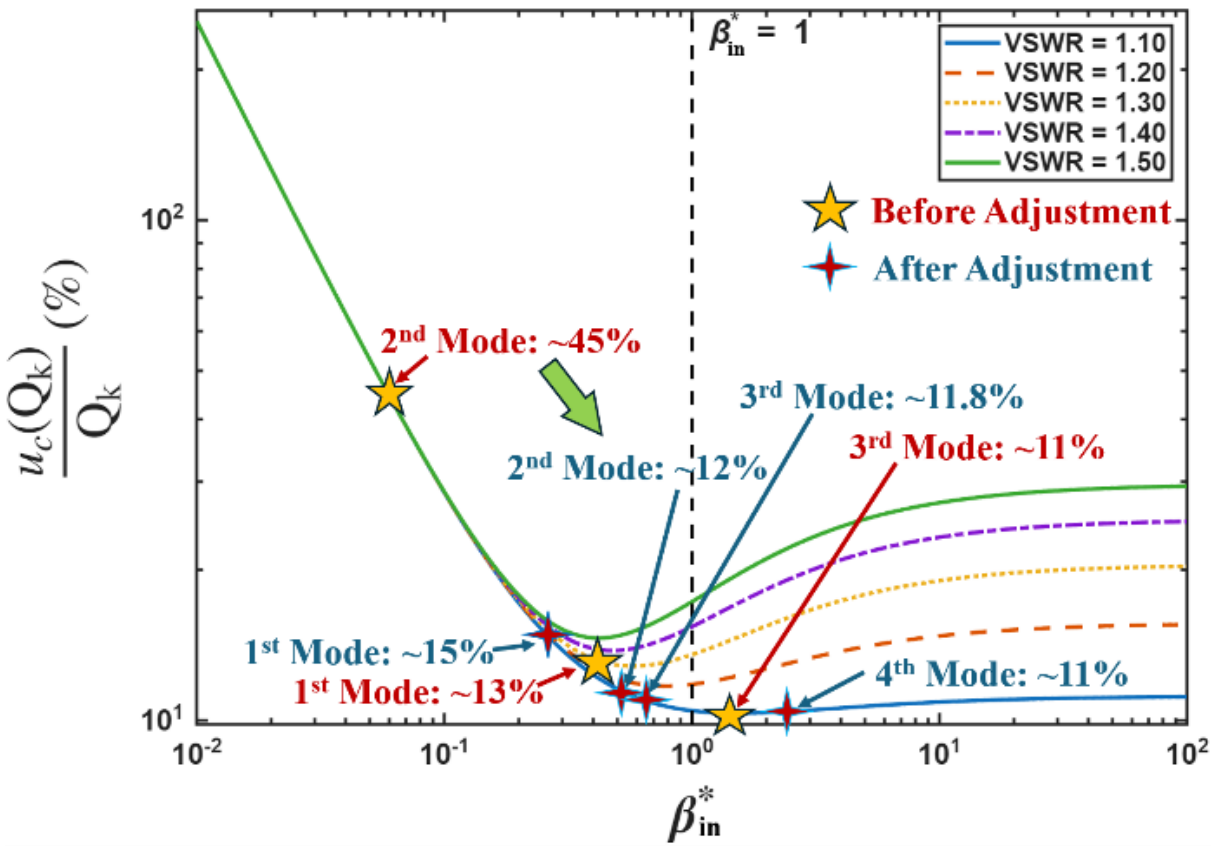

FIG 16. Relative uncertainty of calibrated $Q_\mathrm{k}$ as a function of normalized coupling $\beta_{\mathrm{in}}^*$ for different VSWR values.

This coupling dependence is particularly important for the JLab QPR because the quadrupole modes do not have comparable intrinsic quality factors. The 2nd and 4th quadrupole modes exhibit substantially lower $Q_0$, on the order of $10^6$, which is attributed to additional RF loss of the middle flange, as suggested by the field distributions in FIG 2 and the hardware configuration shown in FIG 3(a). In contrast, the 1st and 3rd quadrupole modes have $Q_0$ values on the order of $10^8$. Because the intrinsic quality factors differ by approximately two orders of magnitude, it is difficult to match all four modes using a single fixed input-coupler configuration. This behavior is reflected in the initial commissioning measurements, where the relative uncertainty in $Q_\mathrm{k}$ for the 2nd mode reached approximately 45%.

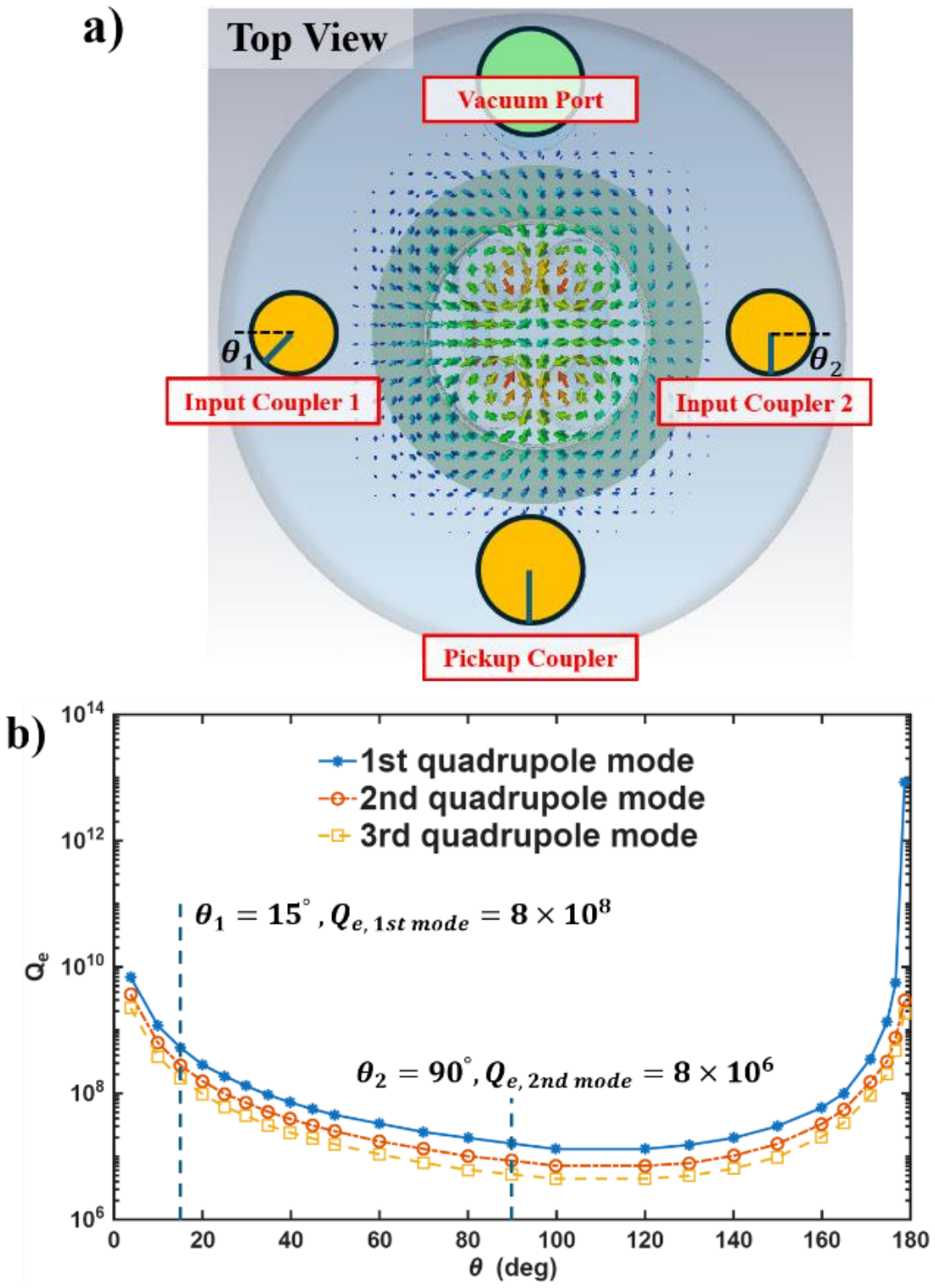

FIG 17. Two-input-coupler scheme for mode-dependent coupling control. (a) Top-view definition of the coupler angles $\theta_1$ and $\theta_2$. (b) CST-simulated $Q_e$ versus coupler angle. The selected angles, $\theta_1 = 15^\circ$ and $\theta_2 = 90^\circ$, match the high-$Q_0$ and low-$Q_0$ modes, respectively.

To mitigate this limitation, a two-port input-coupling scheme was adopted. The transmission port associated with $P_T$ in FIG. 6 was repurposed as the second input-coupler port. The coupling of this second input coupler was adjusted specifically to improve the matching of the 2nd and 4th modes, as shown in FIG 17(a). Based on the simulated dependence of $Q_e$ on the coupler angle, shown in FIG 17(b), coupler 1 was set to $\theta_1 = 15^\circ$, corresponding to $Q_{e,in1} = 8\times10^8$, while coupler 2 was set to $\theta_2 = 90^\circ$, corresponding to $Q_{e,in2} = 8\times10^6$. After this adjustment, the relative uncertainty in $Q_\mathrm{k}$ for the 2nd mode was reduced from approximately 45% to approximately 12%, and that for the 4th mode reached a relative uncertainty of approximately 11%. Meanwhile, the uncertainties for the 1st and 3rd modes remained close to their original levels. During the 1st and 3rd mode tests, the input coupler 2 acted as a strong over-coupled port; therefore, a stub tuner was connected to the coupler port on the top-plate to provide highly reflective terminator to reflect the coupled power back to the resonator. Although this configuration can reduce the effective $Q_0$ of the 1st and 3rd modes, it does not significantly increase the propagated measurement uncertainty. Moreover, because $R_s$ is determined calorimetrically from RF–DC thermal substitution rather than from the resonator effective $Q_0$, a moderate reduction of the resonator effective $Q_0$ does not directly affect the extracted sample surface resistance.

### C. $B_{\mathrm{pk}}$ and $R_s$ Error Propagation

The Jacobian in Eq. (17) gives the sensitivity coefficients of $B_{\mathrm{pk}}$ to each input, Eq. (18) gives the combined standard uncertainty, and Eq. (19) gives the corresponding relative form, which contains two contributions: the uncertainty propagated from $Q_k$ which is discussed in Sec. IV B, and the uncertainty associated with the pickup power $P_\mathrm{k}$ which is summarized in Ref. [36].

$$X_{B_{\mathrm{pk}}} = [Q_\mathrm{k}, P_\mathrm{k}]^T, \quad (16)$$

$$J_{B_{\mathrm{pk}}} = \left[\frac{\partial B_{pk}}{\partial Q_k} \;\; \frac{\partial B_{\mathrm{pk}}}{\partial P_\mathrm{k}}\right] = \left[\frac{B_{\mathrm{pk}}}{2Q_\mathrm{k}} \;\; \frac{B_{\mathrm{pk}}}{2P_\mathrm{k}}\right], \quad (17)$$

$$u_c^2(B_{\mathrm{pk}}) = J_{B_{\mathrm{pk}}} U_{X_{B_{\mathrm{pk}}}} J_{B_{\mathrm{pk}}}^T = \sum_i \left(\frac{\partial B_{\mathrm{pk}}}{\partial X_{B_{\mathrm{pk}},i}}\right)^2 u_c^2\left(X_{B_{\mathrm{pk}},i}\right), \quad (18)$$

$$\left(\frac{u_c(B_{\mathrm{pk}})}{B_{\mathrm{pk}}}\right)^2 = \left(\frac{u_c(Q_\mathrm{k})}{2Q_\mathrm{k}}\right)^2 + \left(\frac{u_c(P_\mathrm{k})}{2P_\mathrm{k}}\right)^2. \quad (19)$$

For the subsequent uncertainty and resolution analysis, a conservative upper-bound relative standard uncertainty of $u(Q_\mathrm{k})/Q_\mathrm{k} = 15\%$ was adopted based on the calibrated-$Q_\mathrm{k}$ uncertainty results shown in FIG 16. The relative standard uncertainty of the pickup-power measurement was taken as $u(P_\mathrm{k})/P_\mathrm{k} = 7\,\%$. Assuming these two contributions are statistically independent, Eq. (19) gives

$$\frac{u_c(B_{\mathrm{pk}})}{B_{\mathrm{pk}}} = \frac{1}{2}\sqrt{\left(\frac{u(Q_\mathrm{k})}{Q_\mathrm{k}}\right)^2 + \left(\frac{u(P_\mathrm{k})}{P_\mathrm{k}}\right)^2} = 8.3\%.$$

Similarly, the relative uncertainty in $R_\mathrm{s}$ in Eq. (23) can be derived from Eqs. (20)-(22). It is governed primarily by two parts: the uncertainty associated with $B_{\mathrm{pk}}$ and the uncertainty associated with the thermal power measurement.

$$X_{R_s} = [B_{\mathrm{pk}} \;\; I_{\mathrm{DC1}} \;\; V_{\mathrm{DC1}} \;\; I_{\mathrm{DC2}} \;\; V_{\mathrm{DC2}}]^T, \quad (20)$$

$$J_{R_s} = \left[\frac{\partial R_s}{\partial B_{\mathrm{pk}}} \;\; \frac{\partial R_s}{\partial I_{\mathrm{DC1}}} \;\; \frac{\partial R_s}{\partial V_{\mathrm{DC1}}} \;\; \frac{\partial R_s}{\partial I_{\mathrm{DC2}}} \;\; \frac{\partial R_s}{\partial V_{\mathrm{DC2}}}\right]$$

$$= \left[ -\frac{2R_s}{B_{pk}} \quad \frac{A\,V_{DC1}}{B_{pk}^2} \quad \frac{A\,I_{DC1}}{B_{pk}^2} \quad -\frac{A\,V_{DC2}}{B_{pk}^2} \quad -\frac{A\,I_{DC2}}{B_{pk}^2} \right], \tag{21}$$

where $A = 2c_1\mu_0^2$.

$$u_c^2(R_s) = J_{R_s}\,U_{R_s}\,J_{R_s}^T = \sum_i \left(\frac{\partial R_s}{\partial X_{R_s,i}}\right)^2 u_c^2(X_{R_s,i}), \tag{22}$$

$$\left(\frac{u_c(R_s)}{R_s}\right)^2 = \left(2\frac{u_c(B_{pk})}{B_{pk}}\right)^2 + \frac{V_{DC1}^2 u_c^2(I_{DC1}) + I_{DC1}^2 u_c^2(V_{DC1}) + V_{DC2}^2 u_c^2(I_{DC2}) + I_{DC2}^2 u_c^2(V_{DC2})}{(I_{DC1}V_{DC1} - I_{DC2}V_{DC2})^2}. \tag{23}$$

Eq.(23) also can be rewritten as Eq. (24).

$$\left(\frac{u_c(R_s)}{R_s}\right)^2 = \left(2\frac{u_c(B_{pk})}{B_{pk}}\right)^2 + \frac{\left(\frac{u_c(P_{DC1})}{P_{DC1}}\right)^2 + r^2\left(\frac{u_c(P_{DC2})}{P_{DC2}}\right)^2}{(1-r)^2}, \tag{24}$$

where $r = \frac{P_{DC2}}{P_{DC1}}$. As $r$ approaches 1, the system enters a pathological region because the second term in Eq. (24) begins to dominate the total uncertainty. Once $r > 0.9$, the total relative uncertainty rises rapidly. Using the conservative upper-bound relative standard uncertainty of 8.3% for $B_{pk}$ and 0.5% for each DC-heater power measurement, Eq. (24) gives $u_c(R_s)/R_s = 18\,\%$ at $r = 0.9$. This implies that measurements performed at low $B_{pk}$, which correspond to $r$ values close to unity, are generally associated with large uncertainty in the extracted $R_s$.

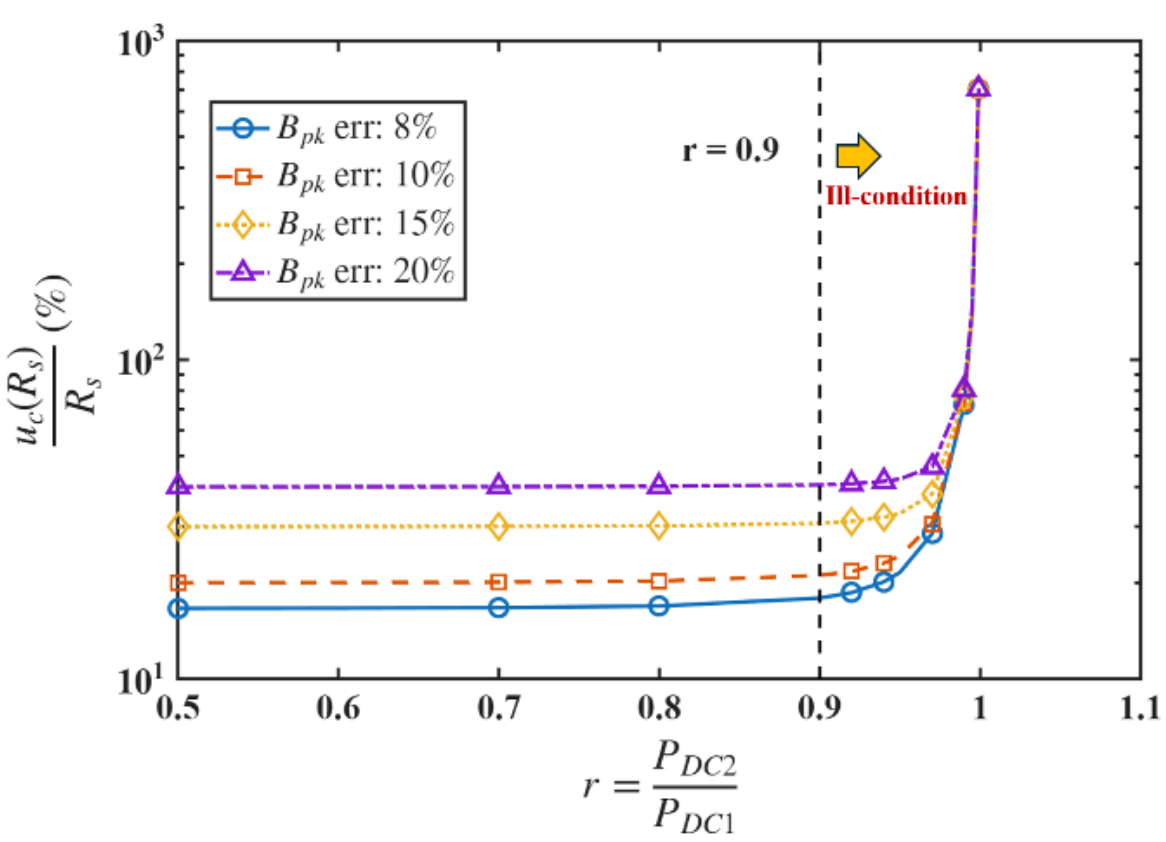

FIG 18. Relative uncertainty of $R_s$ as a function of $r = P_{DC2}/P_{DC1}$ for different $B_{pk}$ uncertainty levels.

FIG 18 shows that the ill-conditioned region with $r > 0.9$ corresponds primarily to measurements performed in the low-field range, approximately 5–10 mT. In this region, the resulting error bars of the extracted $R_s$ are typically greater than 30%. Therefore, the low-field operating regime is not only limited by signal level but also by poor uncertainty performance resulting from the small differences between $P_{DC1}$ and $P_{DC2}$.

## V. RESOLUTION ESTIMATION

The practical resolution of the JLab QPR surface-resistance measurement is governed by two limits. The first is the minimum thermal power resolvable by the RF–DC substitution system. The second is the propagated uncertainty of the extracted surface resistance once the RF-induced heating signal is detectable. The thermal detectability limit is evaluated below from the measured no-RF heater responses of the bulk-Nb and $Nb_3Sn$-Ta-Cu samples and subsequently converted into a minimum detectable surface resistance.

For each sample $m$, the measured RF-off heater power $P_{DC1}$ was fitted using

$$P_{DC1,m}(T, T_{bath}) = A_m(T - T_{bath}) + B_m(T^2 - T_{bath}^2), \tag{25}$$

where $m$ denotes either the bulk-Nb or $Nb_3Sn$-Ta-Cu sample. The Nb coefficients were obtained from a joint fit to the $T_{bath} = 1.6$ and 2.0 K datasets, whereas the $Nb_3Sn$-Ta-Cu coefficients were obtained from a separate fit $T_{bath} = 2.0$ K dataset. Because these fits are derived from the measured $P_{DC1}(T)$ responses of the assembled samples, they represent the effective thermal behavior of the actual heater–sample systems, including thermal contact resistance, heat transport through substrate and interfaces, mounting conditions, and temperature-dependent thermal conductance.

For a temperature-readout resolution $\Delta T_{min}$, the corresponding Cernox-equivalent minimum resolvable heater- power increment is

$$P_{min,T,m}(T) \cong \left|\frac{\partial P_{DC1,m}}{\partial T}\right| \Delta T_{min} = |A_m + 2B_m T|\,\Delta T_{min}. \tag{26}$$

A representative low-temperature readout resolution of $\Delta T_{min} = 0.1$ mK was adopted. The Lake Shore Model 350 specifies a temperature-equivalent measurement resolution of 89 μK for a CX-1050 Cernox sensor at 4.2 K with 10 mV excitation [38]. The heater voltage and current were measured independently using two Keithley 2000 6½-digit digital multimeters [35]. For the measurement ranges used in this work, their voltage and current readout resolutions are sufficiently fine to resolve power changes below the $\Delta P_{DC}$ levels considered here and therefore do not limit the thermal-power detectability.

A second limit arises from the finite current step $\Delta I_{\min}$ of the heater output. For a heater resistance $R_h$, the corresponding minimum power increment is

$$P_{\min,I,m}(T) = |[I_m(T) + \Delta I_{\min}]^2 R_h - I_m^2(T) R_h|,$$

$$\text{and } I_m(T) = \sqrt{\frac{P_{\text{DC1},m}(T)}{R_h}}, \tag{27}$$

where $R_h = 50\,\Omega$ and $\Delta I_{\min} = 10\,\mu\text{A}$. FIG 19 compares the minimum detectable thermal-power limits obtained for the bulk-Nb and $Nb_3Sn$-Ta-Cu samples. Over the overlapping temperature range, the bulk-Nb sample exhibits the larger minimum resolvable power increment, $\Delta P_{\min}$, and therefore represents the upper-bound estimate of the thermal-power resolution of the QPR system. Accordingly, the system-level detection limit used in the subsequent resolution analysis is defined from the bulk-Nb result as

$$P_{min}(T) \equiv P_{\min,Nb}(T) = \max\left[P_{\min,T,Nb}(T), P_{\min,I,Nb}(T)\right]. \tag{28}$$

For the Nb sample, the current-step contribution was evaluated for both bath-temperature conditions included in the thermal calibration, and the larger value was retained. The lower $\Delta P_{\min}$ obtained for the Cu-substrate $Nb_3Sn$-Ta-Cu sample indicates that, for the sample assemblies evaluated in this work, Cu-based samples provide better thermal-power resolution than the conservative Nb-based limit.

Over the evaluated temperature range, the conservative Nb-based $P_{\min}(T)$ envelope shown in FIG 19 was represented by $P_{\min}(T) = -0.0268 + 0.0204T$ mW. This empirical relation was obtained from the measured bulk-Nb substrate $P_{\text{DC1}}(T)$ response through Eqs. (26)-(28) and was used as the conservative system-level input for all four quadrupole modes. Substituting this $P_{\min}(T)$ for $\Delta P_{\text{DC}}$ in Eq. (4) gives

$$R_{s,\min}(T, B_{\text{pk}}) = \frac{2c_1\mu_0^2 P_{\min}(T)}{B_{\text{pk}}^2}. \tag{29}$$

Eq. (29) defines the smallest surface resistance that produces a thermally resolvable RF-heating signal for a given sample temperature and peak magnetic field. Because $R_{s,\min} \propto B_{\text{pk}}^{-2}$, increasing $B_{\text{pk}}$ lowers the thermal detectability floor until the accessible field is constrained by the heater-power budget. FIG 20 shows the resulting minimum detectable surface resistance over the commissioned operating range for each quadrupole mode, using its corresponding $c_1$ value.

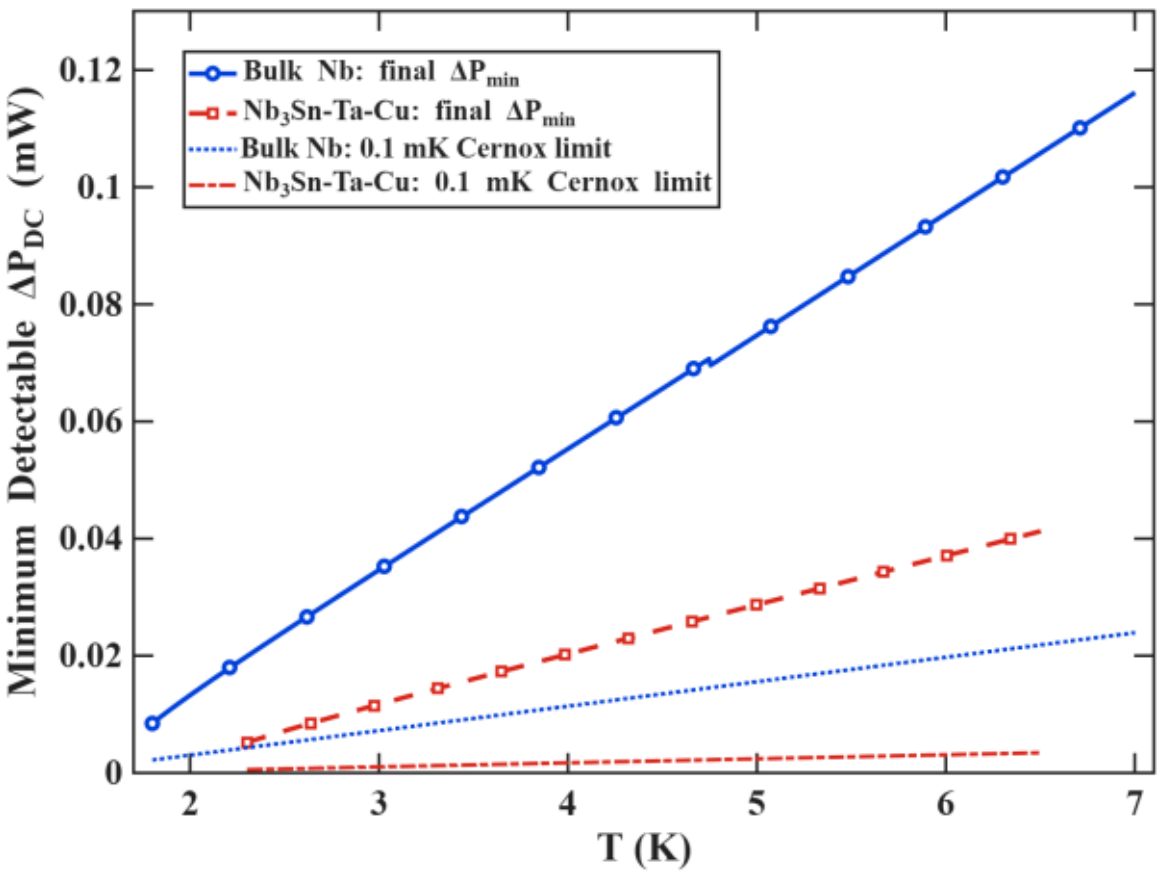

FIG 19. Minimum detectable thermal power for the bulk-Nb and $Nb_3Sn$-Ta-Cu samples. The Cernox-equivalent curves are obtained from the measured $P_{\text{DC1}}(T)$ responses using a temperature resolution of 0.1 mK. The final detection limits are defined as the larger of the Cernox temperature-readout and heater-current-step limits.

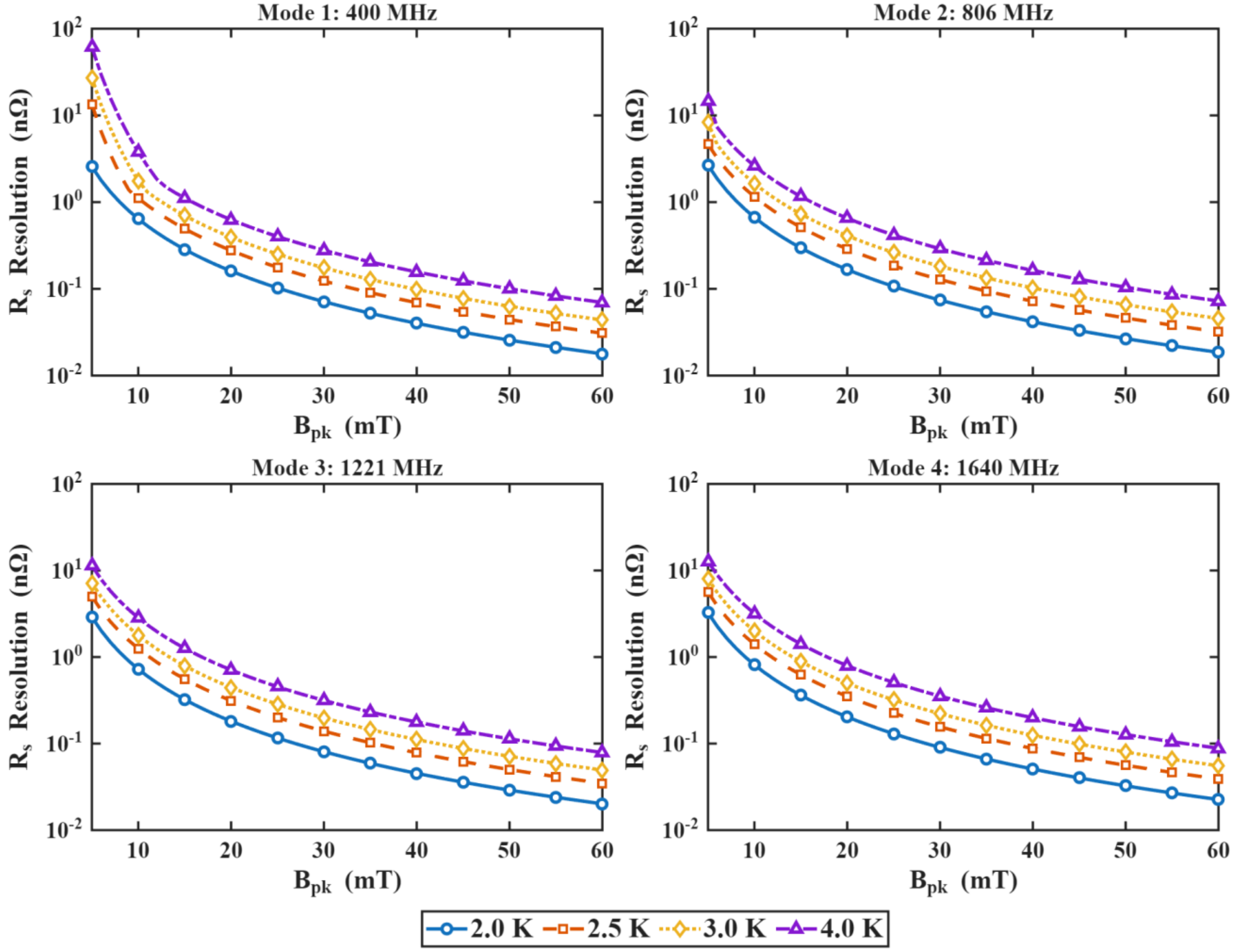

FIG 20**.** Usable surface-resistance resolution of the JLab QPR at 95% confidence level for 400, 806, 1221, and 1640 MHz for sample temperatures of 2.0, 2.5, 3.0, and 4.0 K. The resolution is defined by Eq. (31) as the larger of the minimum thermally detectable surface resistance and the uncertainty-based resolution at the 95% confidence level.

The thermal detectability floor alone does not define the usable surface-resistance resolution. Once the RF-induced heating is detectable, the 95% confidence-level resolution is determined by the combined standard uncertainty derived in Sec. IV:

$$\mathrm{Resolution}_{95\%}(T, B_{\mathrm{pk}}) = 1.96\, u_c(R_s)|_{R_s=R_{s,min}} = 1.96\, R_{s,\min}(T, B_{\mathrm{pk}}) \left.\frac{u_c(R_s)}{R_s}\right|_{R_s=R_{s,min}}, \quad (30)$$

where $u_c$ is the combined standard uncertainty. For the resolution calculation, the large-effective-degrees-of-freedom limit was adopted, for which the two-sided 95% coverage factor approaches 1.96. Accordingly, a coverage factor of $k = 1.96$ was used throughout the resolution analysis. The uncertainty includes the propagated contributions from the calibration, pickup-power measurement, and RF–DC thermal-substitution power measurement.

The experimentally usable resolution must satisfy both the thermal-detectability and confidence-level uncertainty requirements. It is therefore defined as

$$\mathrm{R_s\ Resolution}\,(T, B_{\mathrm{pk}}) = \max[\mathrm{Resolution}_{min}(T, B_{\mathrm{pk}}),\ \mathrm{Resolution}_{95\%}(T, B_{\mathrm{pk}})]. \quad (31)$$

FIG 20 shows the surface-resistance resolution at the 95% confidence level. At low $B_{\mathrm{pk}}$, the small RF-induced heating signal produces a large relative uncertainty in $\Delta P_{\mathrm{DC}}$, and the resolution deteriorates rapidly. Increasing $B_{\mathrm{pk}}$ improves both thermal detectability and uncertainty performance, while the maximum accessible field is ultimately limited by the available heater-power budget.

Analogous to the $R_s$ resolution analysis, the $B_{\mathrm{pk}}$ resolution was evaluated using Eq. (2). The minimum detectable pickup power specified for the RF power meter was converted to the corresponding minimum $B_{\mathrm{pk}}$, while the $Q_{\mathrm{k}}$ uncertainty obtained from FIG 16 and the pickup-power uncertainty ~7 % were propagated to determine the uncertainty-based contribution at the 95% confidence level. The resulting $B_{\mathrm{pk}}$ resolutions are approximately 1.35, 0.91, 0.76, and 0.64 mT at 400, 806, 1221, and 1640 MHz, respectively. Therefore, the most conservative $B_{\mathrm{pk}}$ resolution of the commissioned system is approximately 1.35 mT.

## VI. SUMMARY OF COMMISSIONING

The commissioned JLab QPR supports calibrated SRF sample measurements at four quadrupole-mode frequencies: 400, 806, 1221, and 1640 MHz. The sample temperature can be varied from ~1.8 K to near the superconducting transition temperature, while the accessible peak surface magnetic field extends from approximately 5 mT to a sample- and temperature-dependent heater-power-budget limit, primarily determined by thermal properties such as the thermal conductivity of the sample substrate. A maximum field of 60 mT was demonstrated for the bulk-Nb sample at 400 MHz and 4 K. The combined relative standard uncertainties are 8.3% for $B_{\mathrm{pk}}$and below 18% for $R_s$ when $r < 0.9$ . The most conservative 95% confidence-level $B_{\mathrm{pk}}$ resolution is approximately 1.35 mT at 400 MHz, while the $R_s$ resolution is below 1 nΩ at 10 mT and 2 K. The commissioned operating specifications are summarized in Table V.

Table V. Commissioned operating specifications of the JLab QPR.

| Quantity | Unit | Specification | Comment |
|---|---|---|---|
| Operating Frequencies | MHz | 400, 806, 1221, 1640 | Four quadrupole modes |
| Sample-temperature range | K | 1.8 - $T_c$ | Depends on the sample material |
| $B_{pk}$ range | mT | 5 mT to Heater-power-budget limits | 60 mT demonstrated for the Nb sample |
| $B_{pk}$ combined relative standard uncertainty | % | 8.3 | Worst-case value |
| $R_s$ combined relative standard uncertainty | % | ≤18 | For $r < 0.9$ |
| $B_{pk}$ resolution | mT | ≤1.35 | Worst case at 400MHz; 95% confidence level |
| $R_s$ resolution | nΩ | <1 | At 10 mT, 2 K; 95% confidence level |

## VII. CONCLUSIONS

The Jefferson Lab QPR has been designed, commissioned, and validated as a calibrated platform for SRF material characterization. Optimization of the pole-shoe geometry resolved the mode-overlap limitation and enabled four usable quadrupole modes at 400, 806, 1221, and 1640 MHz. Commissioning measurements on bulk Nb and $Nb_3Sn$-Ta-Cu samples demonstrated RF–DC thermal-substitution measurements over broad ranges of field, temperature, and frequency. The extracted superconducting energy-gap parameters were consistent with established material values and with those obtained from the corresponding single-cell cavity measurements.

A comprehensive uncertainty-propagation framework was developed for the one-time $Q_{\mathrm{k}}$ calibration, $B_{\mathrm{pk}}$ determination, thermal-substitution power measurement, and evaluation of measurement resolution at a specified confidence level. The analysis identifies the coupling condition, heater-power budget, and weak low-field thermal signal as the principal practical limitations. The commissioned system provides conservative $B_{\mathrm{pk}}$ resolution of approximately 1.35 mT at the 95% confidence level and an $R_{\mathrm{s}}$ resolution below 1 nΩ at 10 mT and 2 K. These results establish the JLab QPR as a calibrated multi-frequency platform for screening and investigating Nb, $Nb_3Sn$, and other advanced thin-film structures for SRF cavity applications.

## ACKNOWLEDGMENTS

The authors thank Piotr Putek for discussions on the electromagnetic design optimization, Jiquan Guo for reviewing the electromagnetic simulations, Sarra Overstreet for component fabrication and procurement, Christiana Wilson for assistance with LabVIEW programming, the Jefferson Lab cavity-processing team for preparing the QPR for vertical testing, and the Jefferson Lab VTA team for cryogenic testing and RF-system support. This material is based upon work supported by the U.S. Department of Energy, Office of Science, Office of Nuclear Physics, under Contract No. 89243126CSC000213.

## DATA AVAILABILITY

The data that support the findings of this study are available from the corresponding author upon reasonable request.

## APPENDIX A: RF POWER BALANCE FOR TWO- AND THREE-COUPLER CONFIGURATIONS

This appendix derives the RF power-balance expressions used to compare the two- and three-coupler QPR configurations. All quantities are evaluated at resonance. The stored energy is related to the peak magnetic field on the sample by

$$U = \frac{B_{\rm pk}^2}{c_2}, \tag{A1}$$

where $c_2 = B_{\rm pk,samp}^2/U$ is obtained from electromagnetic simulation.

For a resonator driven through the input port, define the effective non-input quality factor as

$$\frac{1}{Q_{\rm eff}} = \frac{1}{Q_0} + \frac{1}{Q_{\rm k}} + \frac{1}{Q_{\rm T}}, \tag{A2}$$

where $Q_0$ is the intrinsic quality factor, and $Q_{\rm k}$ and $Q_{\rm T}$ are the external quality factors of the pickup and transmission ports, respectively. The corresponding effective input coupling factor is

$$\beta_{\rm in}^* = \frac{Q_{\rm eff}}{Q_{\rm in}}, \tag{A3}$$

where $Q_{\rm in}$ is the external quality factor of the input port. The on-resonance forward power required to maintain the stored energy $U$ is

$$P_F = \frac{\left(1+\beta_{in}^*\right)^2}{4\beta_{in}^*}\frac{\omega_0 U}{Q_{\rm eff}} = \frac{\left(1+\beta_{in}^*\right)^2}{4\beta_{in}^*}\left(\frac{\omega_0 U}{Q_0} + \frac{\omega_0 U}{Q_k} + \frac{\omega_0 U}{Q_T}\right). \tag{A4}$$

The corresponding reflected power is

$$P_R = P_F\left(\frac{1-\beta_{in}^*}{1+\beta_{in}^*}\right)^2. \tag{A5}$$

Eqs. (A4) and (A5) apply to both configurations. The two-coupler case is obtained by setting $1/Q_T = 0$, whereas the three-coupler case retains the additional transmission-port loss term.

For the three-coupler comparison shown in FIG 7, the condition $Q_T = Q_{in}$ was imposed. Defining $A \equiv 1/Q_0 + 1/Q_k$, the effective coupling factor becomes

$$\beta_{\rm in,3C}^* = \frac{1}{1+AQ_{in}},\ \ 0 < \beta_{in,3C}^* < 1. \tag{A6}$$

Using this relation, the required forward power can be written as

$$P_{F,3\rm C} = \frac{\left(1+\beta_{\rm in,3C}^*\right)^2}{4\beta_{\rm in,3C}^*\left(1-\beta_{\rm in,3C}^*\right)}\ \omega_0 UA. \tag{A7}$$

Eq. (A7) makes the three-coupler power trade-off explicit. When the original two-coupler configuration is over-coupled, the transmission port can reduce the reflected power by shifting the effective input coupling toward critical coupling. However, the transmission port also extracts RF power from the resonator and therefore increases the required forward power. If the original configuration is already under-coupled, the additional loading moves the effective coupling farther below unity and increases rather than reduces the reflected power.

## APPENDIX B: CLOSED-FORM CALIBRATION OF THE PICKUP EXTERNAL QUALITY FACTOR

This appendix derives the closed-form expression for the pickup-port external quality factor $Q_k$ used in the one-time decay calibration. All RF powers are referred to the cavity reference plane after cable-loss correction.

Using the effective non-input quality factor and effective input coupling factor defined in Eqs. (A2) and (A3), together with the definition of the loaded quality factor, one obtains

$$Q_{\rm eff} = (1+\beta_{\rm in}^*)Q_L. \tag{B1}$$

Combining the forward-power balance in Eq. (A4), the reflected-power relation in Eq. (A5), Eq. (B1), and the decay relation $Q_L = \omega_0\tau$ from Eq. (9) gives

$$Q_{\rm k} = (1+\beta_{\rm in}^*)\frac{P_F - P_R}{P_k}\omega_0\tau, \tag{B2}$$

where $\tau$ is the stored-energy decay time constant. Defining the measured on-resonance reflection magnitude as

$$\rho = \sqrt{\frac{P_R}{P_F}},$$

the reflected-power relation in Eq. (A5) gives

$$\beta_{\rm in}^* = \frac{1\pm\rho}{1\mp\rho}, \tag{B3}$$

where the upper signs apply to the over-coupled condition $\beta_{in}^* > 1$, and the lower signs apply to the under-coupled condition, $\beta_{in}^* < 1$.

Substituting Eq. (B3) into Eq. (B2) and using $P_F - P_R = P_F(1 - \rho^2)$ gives

$$Q_\mathrm{k} = \frac{2\sqrt{P_F}\left(\sqrt{P_F} \pm \sqrt{P_R}\right)\omega_0\tau}{P_\mathrm{k}}, \tag{B4}$$

where the $+$ sign applies to the over-coupled condition and the $-$ sign applies to the under-coupled condition. At critical coupling, $P_\mathrm{R} = 0$, and the two branches coincide.